\documentclass[dvips,aos,preprint]{imsart}
\usepackage{mathrsfs}

\RequirePackage[OT1]{fontenc}
\usepackage[authoryear,round,compress,sort]{natbib}   % reference without numbers
\usepackage{amsmath,amssymb,amsfonts}
\usepackage{amsbsy, amsthm,epsfig}
\usepackage{graphicx,rotating, tabularx, booktabs}
\usepackage{color}
\usepackage{pdflscape}
\usepackage{subfigure}

\startlocaldefs
\numberwithin{equation}{section}
\theoremstyle{plain}

\endlocaldefs
\def\be{\begin{equation}}
\def\ee{\end{equation}}
\def\bea{\begin{eqnarray}}
\def\eea{\end{eqnarray}}
\def\bd{\begin{displaymath}}
\def\ed{\end{displaymath}}
\def\bda{\begin{eqnarray*}}
\def\eda{\end{eqnarray*}}
\def\bsm{\begin{small}}
\def\esm{\end{small}}

\def\ha1{\hat \beta_1}

\def\T{{ \mathrm{\scriptscriptstyle T} }}
\def\AS{A\"{\i}t-Sahalia}

\def\bsc{\begin{scriptsize}}
\def\esc{\end{scriptsize}}

\usepackage{amssymb}
\def\be{\begin{equation}}
\def\ee{\end{equation}}
\def\bea{\begin{eqnarray}}
\def\eea{\end{eqnarray}}
\def\bd{\begin{displaymath}}
\def\ed{\end{displaymath}}
\def\bda{\begin{eqnarray*}}
\def\eda{\end{eqnarray*}}

\def\ha1{\hat \beta_1}

\def\AS{A\"{\i}t-Sahalia}

\def\bsc{\begin{scriptsize}}
\def\esc{\end{scriptsize}}

\newtheorem{theorem}{Theorem}

\newtheorem{proposition}{Proposition}

\theoremstyle{definition}
\newtheorem{as}{Condition}

\newtheorem{example}{Example}
\newtheorem{remark}{Remark}

\newcommand{\calM}{\mathcal{M}}
\newcommand{\calF}{\mathcal{F}}
\newcommand{\calB}{\mathcal{B}}

\newcommand{\wh}{\widehat}

\def\sep{\mathrm{sep}}
\def\thre{\mathrm{thre}}

\makeatletter
\newcommand{\figcaption}{\def\@captype{figure}\caption}
\newcommand{\tabcaption}{\def\@captype{table}\caption}
\makeatother

\newcommand{\cor}{{\rm Corr}}
\newcommand{\cov}{{\rm Cov}}
\newcommand{\diag}{{\rm diag}}

\newcommand{\tr}{\mbox{tr}}
\newcommand{\var}{\mbox{Var}}

\def\de{\delta}

\def\la{\lambda}

\newcommand{\ve}{{\varepsilon}}

\newcommand{\bA}{{\mathbf A}}
\newcommand{\bB}{{\mathbf B}}

\newcommand{\bH}{{\mathbf H}}
\newcommand{\bI}{{\mathbf I}}

\newcommand{\bS}{{\mathbf S}}

\newcommand{\bV}{{\mathbf V}}
\newcommand{\bW}{{\mathbf W}}

\newcommand{\bu}{{\mathbf u}}

\newcommand{\bx}{{\mathbf x}}
\newcommand{\by}{{\mathbf y}}
\newcommand{\bz}{{\mathbf z}}

\newcommand{\bSigma}{\boldsymbol{\Sigma}}

\newcommand{\bmu} {\boldsymbol{\mu}}

\newcommand{\bGamma} {\boldsymbol{\Gamma}}

\newcommand{\bD}{{\mathbf D}}
\newcommand{\bzero}{{\mathbf 0}}

\def\JRSSB{{\sl J. Roy. Stat. Soc. B.}}

\def\BKA{{\sl Biometrika}}
\def\JASA{{\sl J. Am. Stat. Assoc.}}

\def\SS{{\sl Statist. Sinica}}

\def\AS{{\sl Ann. Stat.}}
\def\JMA{{\sl J. Multivariate Anal.}}

\def\JOE{{\sl J. Econometrics}}
\def\JTSA{{\sl J. Time Ser. Anal.}}

\renewcommand{\thetable}{ \arabic{table}}

\begin{document}

\begin{frontmatter}
\title{Principal Component Analysis for Second-order Stationary
Vector Time Series}
\runtitle{PCA for Time Series}
\begin{aug}

\author{\fnms{Jinyuan} \snm{Chang}$^{1,}$\thanksref{t1}\ead[label=e1]{changjinyuan@swufe.edu.cn}\ead[label=e2]{guobin@swufe.edu.cn}}
\address{Center of Statistical Research\\
and\\
School of Statistics\\
Southwestern University of Finance\\
~~and Economics\\
Chengdu, Sichuan 611130\\
 China\\
\printead{e1,e2}\\
},
\author{\fnms{Bin} \snm{Guo}$^{1,}$\thanksref{t2}}
 and \author{\fnms{Qiwei} \snm{Yao}$^{2,}$\thanksref{t3}\ead[label=e3]{q.yao@lse.ac.uk}}
\address{Department of Statistics\\
London School of Economics \\
~~and Political Science\\
London, WC2A 2AE\\
 UK\\
\printead{e3}}

\thankstext{t1}{Supported in part by the Fundamental Research Funds for the Central Universities (Grant No.
 JBK150501), NSFC (Grant No. 11501462), the Center of Statistical Research at
SWUFE, and the Joint Lab of Data Science and Business Intelligence at SWUFE.}
\thankstext{t2}{Supported in part by the Fundamental Research Funds for the Central Universities (Grant No.
JBK120509, JBK140507) and the Center of Statistical Research at
SWUFE.}
\thankstext{t3}{Supported in part
by an EPSRC research grant.}
\runauthor{J. CHANG, B. GUO and Q. Yao}

\affiliation{Southwestern University of Finance and Economics$^1$, and London School of Economics and Political Science$^2$}

\end{aug}

\begin{abstract}
We extend the principal component analysis (PCA) to
second-order stationary vector
time series in the sense that  we seek for a contemporaneous linear transformation for
a $p$-variate time series such that the transformed series is segmented
into several lower-dimensional subseries, and those subseries are uncorrelated with
each other both contemporaneously and serially. Therefore those  lower-dimensional series
can be analysed separately as far as the linear dynamic structure is concerned.
Technically it boils down to an eigenanalysis for a positive
definite matrix. When $p$ is large, an additional step is required
to perform a permutation in terms of either maximum
cross-correlations or FDR based on multiple tests. The asymptotic
theory is established for both fixed $p$ and diverging $p$ when the
sample size $n$ tends to infinity. Numerical experiments with both
simulated and real data sets indicate that the proposed method is an
effective initial step in analysing multiple time series data, which
leads to substantial dimension reduction in modelling and
forecasting high-dimensional linear dynamical structures. Unlike PCA
for independent data, there is no guarantee that the required linear
transformation exists. When it does not, the proposed method
provides an approximate segmentation which leads to the advantages
in, for example, forecasting for future values. The method can also
be adapted to segment multiple volatility processes.

\end{abstract}



\end{frontmatter}

\section{Introduction}
Modelling multiple time series, also called vector time series, is
always a challenge, even when the vector dimension $p$ is moderately
large. While most the inference methods and the associated theory
for univariate autoregressive and moving average (ARMA) processes
have found their multivariate counterparts \citep{Lutkepohl_(2006)},
vector autoregressive and moving average (VARMA) models are seldom
used directly in practice when $p\ge 3$. This is partially due to
the lack of identifiability for VARMA models in general. More
fundamentally, those models are overparametrized; leading to flat
likelihood functions which cause innate difficulties in statistical
inference. Therefore finding an effective way to reduce the number
of parameters is particularly felicitous in modelling and
forecasting multiple time series. The urge for doing so is more
pertinent in this modern information age, as it has become
commonplace to access and to analyse  high dimensional time series
data with dimension $p$ in the order of hundreds or more. Big time
series data arise from, among others, panel study for economic and
natural phenomena, social network, healthcare and
public health, financial market, supermarket transactions,
information retrieval and recommender systems.

Available methods to reduce the number of parameters in modelling vector
time series can be divided into two categories: regularization and
dimension reduction. The former imposes some conditions on the structure
of a VARMA model. The latter represents a high-dimensional process in
terms of several lower-dimensional processes. Various regularization
methods have been developed in literature. For example,
\cite{JakemanSteeleYoung_1980} adopted a two stage regression strategy
based on instrumental variables to avoid using moving average explicitly.
Different canonical structures are imposed on VARMA models
[Chapter 3 of \cite{Reinsel_1993}, Chapter 4 of \cite{Tsay_2014}, and
references within].
Structural restrictions are imposed in order to specify and to
estimate some reduced forms of vector autoregressive (VAR) models
%[Chapter 9 of \cite{Lutkepohl_(2006)}, and references within].
[Chapter 9 of \cite{Lutkepohl_(2006)}, and references within].
\cite{Davis2012} proposed a VAR model with sparse coefficient matrices
based on partial spectral coherence. Under different sparsity
assumptions, VAR models have been estimated by LASSO regularization
\citep{ShojaieMichailidis_2010,SongBickel_2011}, or by the Dantzig
selector \citep{HanLiu_2013}. \cite{GuoWangYao_2014} considered  high-dimensional autoregression with banded coefficient matrices.
The dimension reduction methods include the canonical correlation
analysis of \cite{BoxTiao_1977}, the independent components analysis (ICA)
of \cite{BackWeigend_1997}, the principal components analysis (PCA)
of \cite{StockWatson_2002}, the scalar component analysis of
\cite{TiaoTsay_1989} and \cite{Huang_2014}, the dynamic orthogonal
components analysis of \cite{MattesonTsay_2011}. Another popular
approach is to represent multiple time series in terms of a few
latent factors defined in various ways. There is a large body of
literature in this area published in the outlets in statistics,
econometrics and signal processing. An incomplete list of the
publications includes
\cite{Anderson_1963}, \cite{PenaBox_1987}, \cite{Tong_1994}, \cite{Belouchrani_1997},
 \cite{BaiNg_Econometrica_2002}, \cite{Theis_2004}, \cite{StockWatson_2005},
 \cite{Fornietal_2005}, \cite{PanYao_2008},
\cite{LamYaoBathia_Biometrika_2011}, \cite{LamYao_AOS_2012} and
\cite{ChangGuoYao_2013}.

A new dimension reduction method is proposed in this paper. We seek for a
contemporaneous linear transformation such that the transformed series is
segmented into several lower-dimensional subseries, and those subseries
are uncorrelated with each other both contemporaneously and serially.
Therefore they can be modelled or forecasted separately, as far as linear
dependence is concerned. This reduces the number of parameters involved
in depicting linear dynamic structure substantially.
While the basic idea
is not new, which has been explored with various methods
including some aforementioned
references, the method proposed in this paper (i.e. the new PCA for time series)
 is new, simple and effective.
Technically the proposed method boils down to an eigenanalysis for a
positive definite
matrix which is a quadratic function of the cross correlation matrix
function for the observed process. Hence it is easy to implement and the
required computation can be carried out with, for example, an ordinary
personal computer or laptop for the data with dimension $p$ in the order
of thousands.

The method can be viewed as an extension of the standard PCA for
multiple time series, therefore, is abbreviated as TS-PCA. However
the segmented subseries are not guaranteed to exist as those
subseries must not correlate with each other across all
times. This is a marked difference from the standard PCA. The real
data examples in Section \ref{sec5} indicate that it is
often reasonable to assume that the segmentation exists.
Furthermore, when the assumption is invalid, the proposed method
provides some approximate segmentations which ignore some weak
though significant correlations, and those weak correlations are of
little practical use for modelling and forecasting. Thus the
proposed method can be used as an initial step in analysing multiple
time series, which often transforms a multi-dimensional problem into
several lower-dimensional problems. Furthermore the results obtained
for the transformed subseries can be easily transformed back to the
original multiple time series. Illustration with real data examples
indicates clearly the advantages in post-sample forecasting from
using the proposed TS-PCA. The {\sl R}-package {\tt PCA4TS},
available from CRAN project, implements the proposed methodology.

The proposed TS-PCA can be viewed as a version of ICA. In fact our goal is the same in principle
as the ICA using autocovariances presented in Section 18.1 of \cite{Hyvarinenetal_2001}.
However, the nonlinear optimization algorithms presented there are  to
search for a linear transformation such that all the off-diagonal
elements of the autocovariance matrices for the transformed vector time
series are minimized. See also \cite{Tong_1994}, and \cite{Belouchrani_1997}.
To apply those algorithms to our setting, we need to
know exactly the block diagonal structure of autocovariances of
the transformed vector process (i.e. the number of blocks and
the sizes of all the blocks), which is unknown in practice. Furthermore,
our method is simple and fast, and therefore is applicable to high-dimensional
cases. \cite{Cardoso_1998} extends the basic idea of ICA to the so called
multivariate ICA, which requires the transformed random vector to be segmented
into several independent groups with possibly more than one component in
each group. But \cite{Cardoso_1998} does not provide a pertinent
algorithm for multivariate ICA.  Furthermore it does not consider the
dependence across different time lags. TS-PCA is also different from the dynamic PCA proposed
in Chapter 9 of \cite{Brillinger_1981} which decomposes each
component time series as the sum of moving averages of several
uncorrelated white noise processes. In our TS-PCA, no lagged
variables enter the decomposition.

The rest of the paper is organised as follows. The methodology is
spelled out in Section \ref{sec2}. Section \ref{sec3} presents the associated asymptotic
properties of the proposed method. Numerical illustration with real data
are reported in Section \ref{sec5}. Section \ref{se:volatility}
extends the method to segmenting a multiple volatility process into
several lower-dimensional volatility processes. Some final remarks
are given in Section \ref{se:finalremark}. All technical
proofs and numerical illustration with simulated data are relegated to the supplementary material [\cite{CGY_supple}].
%Appendix.
We always use the following notation. For any $m\times k$
matrix $\bH=(h_{i,j})$, let
$
\|\bH \|_2 = \lambda_{\max}^{1/2}(\bH\bH^\T)$ %, $\|\bH \|_\infty =\max_{1\leq i\leq m}\sum_{j=1}^k|h_{i,j}|$
and $\|\bH\|_F=
(\sum_{i=1}^m\sum_{j=1}^kh_{i,j}^2)^{1/2},
$
where $\lambda_{\max}(\bH\bH^\T)$ denotes the largest eigenvalue of $\bH\bH^\T$.

\setcounter{equation}{0}

\section{Methodology}
\label{sec2}

\subsection{Setting and method}
\label{sec21}
Let $\by_t$ be observable $p\times 1$ weakly stationary time series.
We assume that $\by_t$ admits a latent segmentation structure:
\begin{equation} \label{b1}
\by_t = \bA \bx_t,
\end{equation}
where $\bx_t$ is an unobservable $p\times 1$ weakly stationary time
series consisting of $q\; (>1)$ both contemporaneously and serially
uncorrelated subseries, and $\bA$ is an unknown constant matrix.
Hence all the autocovariances of $\bx_t$ are of the same
block-diagonal structure with $q$ blocks.
Denote the segmentation of $\bx_t$ by
\begin{equation} \label{b7}
\bx_t = (
\bx_t^{(1),\T},\ldots,\bx_t^{(q),\T})^{\T}
\end{equation}
with $\cov(\bx_t^{(i)}, \bx_s^{(j)}) ={\bf0}$
for all $t, s$ and $i\ne j$. Therefore
$\bx_t^{(1)}, \ldots, \bx_t^{(q)}$ can be modelled or forecasted
separately as far as their linear dynamic structure is concerned.

\begin{example} \label{ex-temperature}
Before we spell out how to find the segmentation transformation $\bA$ in general,
we consider the monthly temperatures of 7 cities (Nanjing,
Dongtai, Huoshan, Hefei, Shanghai, Anqing and Hangzhou) in Eastern
China from
January 1954 to December 1998. Fig \ref{7cityTemACF}(a) plots
 the cross correlations of these 7 temperature time series.
Both the autocorrelation of each component series and the cross
correlation between any two component series are dominated by the annual
temperature fluctuation; showing the strong  periodicity with the
period 12. Now we apply the linear transformation $\bx_t = \bA^{-1} \by_t$ with
\[
\bA^{-1}= \left(
\begin{array}{rrrrrrr}
0.244& -0.066&   0.019&  -0.050& -0.313& -0.154& 0.200\\
-0.703&  0.324& -0.617&  0.189&  0.633&  0.499& -0.323\\
0.375&  1.544& -1.615&  0.170& -2.266&  0.126& 1.596\\
3.025& -1.381& -0.787& -1.691 &-0.212&  1.188& -0.165\\
-0.197& -1.820& -1.416&  3.269&  0.301& -1.438& 1.299\\
-0.584& -0.354&  0.847& -1.262& -0.218& -0.151& 1.831\\
1.869& -0.742&  0.034&  0.501&  0.492& -2.533& 0.339
\end{array}
\right),
\]
where ${\bf A}$ is determined by the method given in Section 2.
Fig
\ref{7cityTemACF}(b) shows that the first two transformed component
series are significantly correlated both concurrently and serially,
and there are also small but significant correlations in the
(3,\,2)-th panel; indicating the correlations between
the 2nd and the 3rd component series. Apart from these, there is
little significant cross correlation among all the other pairs of
component series. This visual observation suggests to segment the 7
transformed series into 5
uncorrelated groups: \{1,\,2,\,3\}, \{4\}, \{5\}, \{6\} and \{7\}.

\begin{figure}[tbh]
\begin{center}
\subfigure{\includegraphics[scale=0.75]{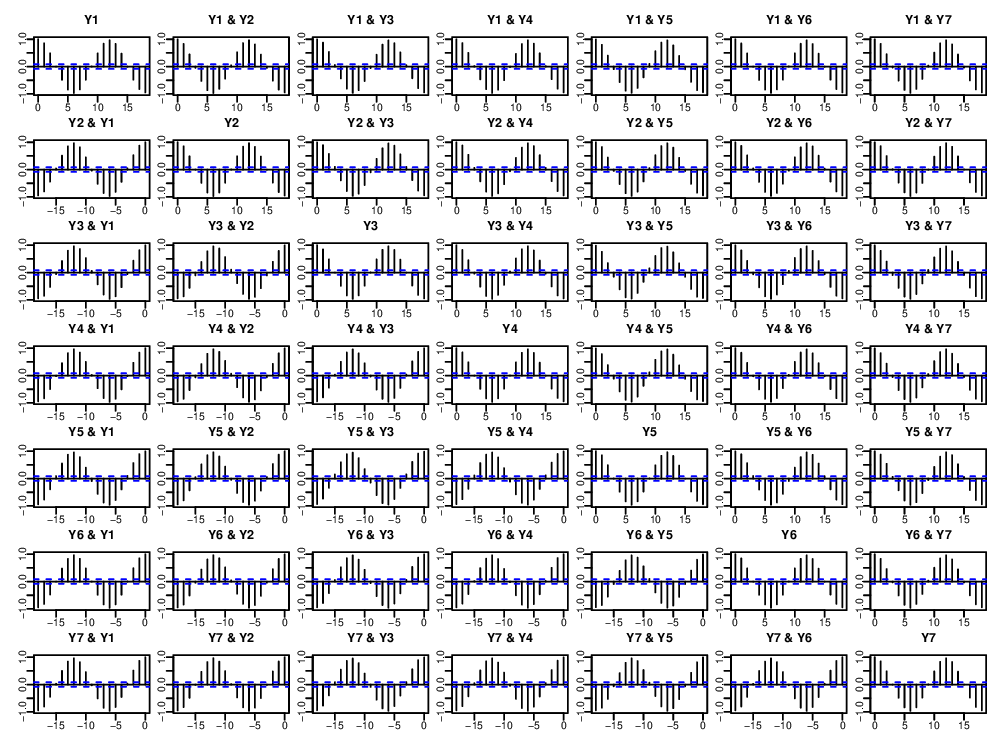}}\\
(a) Cross correlogram of the original $7$ temperature time series.\\
%\subfigure {\includegraphics[scale=0.363]{3exptong1000n5030new}}\\
\subfigure{\includegraphics[scale=0.75]{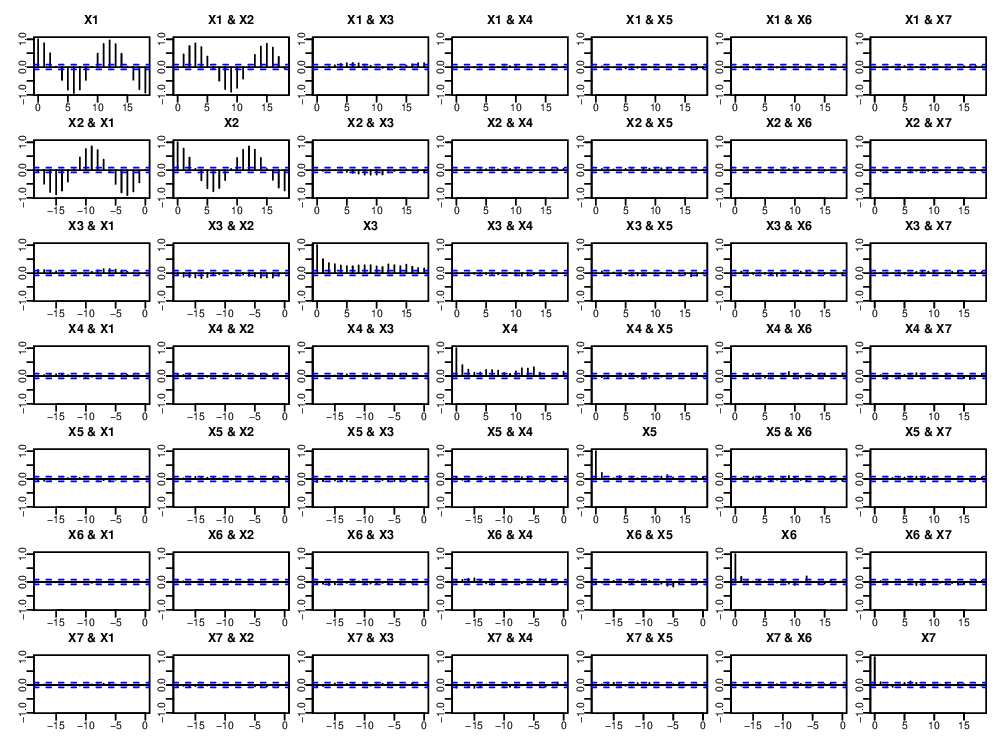}}\\
(b) Cross correlogram of the $7$ transformed component time series.
%\subfigure {\includegraphics[scale=0.363]{3exptong1000n10060}}\\
\setlength{\abovecaptionskip}{0pt}
\end{center}

\caption{Cross correlograms for Example \ref{ex-temperature}.}\label{7cityTemACF}
\end{figure}
\end{example}

This example indicates that the segmentation transformation transfers
the problem of analysing a 7-dimensional time series into
the five lower-dimensional problems: four univariate time series and one
3-dimensional time series. Those five time series can and should be
analysed separately as there are no cross correlations among
them at all time lags. The linear dynamic structure of the original
series is deduced by those of the five transformed series, as
$\cov(\by_{t+k}, \by_t) = \bA \cov(\bx_{t+k}, \bx_t) \bA^\T$.

Now we spell out how to find the segmentation transformation under (\ref{b1}) and
(\ref{b7}).
Without the loss of
generality we may assume
\begin{equation} \label{b2}
\var(\by_t) = \bI_p ~~{\rm and}~~ \var(\bx_t)  = \bI_p,
\end{equation}
where $\bI_p$ denotes the $p\times p$ identity matrix. This first
equation in (\ref{b2})
 amounts to replace $\by_t$ by
$\wh \bV^{-{1/2}}\by_t$ as a preliminary step in practice, where $\wh
\bV$ is a consistent estimator for $\var(\by_t)$.
As both $\bA$ and
$\bx_t$ are unobservable, the second equation in (\ref{b2}) implies that
we view   $(\bA \{ \var(\bx_t)\}^{1/2}, \linebreak \{
\var(\bx_t)\}^{-{1/2}}\bx_t)$ as  $(\bA, \bx_t)$ in (\ref{b1}). More
importantly, the latter perspective will not alter the block-diagonal
structure of the autocovariance matrices of $\bx_t$. Now it follows from
(\ref{b1}) and (\ref{b2}) that
$
\bI_p = \var(\by_t) = \bA \var(\bx_t) \bA^\T = \bA \bA^\T.
$
Thus, $\bA$ in (\ref{b1}) is an orthogonal matrix under (\ref{b2}).

Let $p_j$ be the length of $\bx_t^{(j)}$. %Then $\sum_j p_j =p$.
Write $\bA = (\bA_1, \ldots, \bA_q)$, where $\bA_j$ has $p_j$
columns. Since $\bx_t = \bA^\T \by_t$, it follows from (\ref{b7}) that
\begin{equation} \label{b8}
\bx_t^{(j)} = \bA_j^\T \by_t, \quad j =1, \ldots, q.
\end{equation}
Let $\bH_j$ be any $p_j \times p_j$ orthogonal matrix, and
 $\bH = \diag(\bH_1, \ldots, \bH_q)$.
Then $(\bA, \, \bx_t)$ in (\ref{b1}) can be replaced by $ (\bA \bH,
\, \bH^\T\bx_t)$ while (\ref{b7}) still holds. Hence $\bA$ and
$\bx_t$ are not uniquely identified in (\ref{b1}), even with the additional assumption
(\ref{b2}). In fact under (\ref{b2}),
 only $\calM(\bA_1), \ldots,
\calM(\bA_q)$ are uniquely defined by (\ref{b1}),
where $\calM(\bA_j)$ denotes the linear space spanned by the columns of
$\bA_j$. Consequently,
 $\bGamma_j^\T\by_t$ can be taken as $\bx_t^{(j)}$
for any $p \times p_j$ matrix $\bGamma_j$ as long as
$\bGamma_j^\T\bGamma_j= \bI_{p_j} $ and $\calM(\bGamma_j )
=\calM(\bA_j)$.

To discover the latent segmentation, we need to estimate $\bA =
(\bA_1, \ldots, \bA_q)$, or more precisely, to estimate linear
spaces $\calM(\bA_1) , \ldots, \calM(\bA_q)$. To this end, we
introduce some notation first. For any integer $k$, let
$
\bSigma_y(k) = \cov( \by_{t+k}, \by_t)$ and $\bSigma_x(k) =
\cov( \bx_{t+k}, \bx_t).
$
For a prescribed positive integer $k_0$, define
\begin{align}
\label{Wy} \bW_y & = \sum_{k=0}^{k_0} \bSigma_y(k) \bSigma_y(k)^\T=
\bI_p
+\sum_{k=1}^{k_0} \bSigma_y(k) \bSigma_y(k)^\T, \\
\bW_x  &= \sum_{k=0}^{k_0} \bSigma_x(k) \bSigma_x(k)^\T = \bI_p
+\sum_{k=1}^{k_0} \bSigma_x(k) \bSigma_x(k)^\T. \nonumber
\end{align}
Then both $\bSigma_x(k)$ and $\bW_x$ are block-diagonal, and
\begin{equation} \label{b3}
\bW_y = \bA \bW_x \bA^\T.
\end{equation}
Note that both $\bW_y$ and $\bW_x$ are positive definite matrices.
Let
\begin{equation} \label{b5}
\bW_x \bGamma_x = \bGamma_x \bD,
\end{equation}
i.e. $\bGamma_x$ is a $p\times p$ orthogonal matrix with the columns
being the orthonormal eigenvectors of $\bW_x $, and $\bD$ is a
diagonal matrix with the corresponding eigenvalues as the elements
on the main diagonal. Then (\ref{b3}) implies that
$
\bW_y \bA \bGamma_x = \bA \bGamma_x \bD.
$
Hence the columns of $\bGamma_y \equiv  \bA \bGamma_x $ are the
orthonormal eigenvectors of $\bW_y$. Consequently,
\begin{equation} \label{b6}
\bGamma_y^\T \by_t =  \bGamma_x^\T \bA^\T \by_t = \bGamma_x^\T \bx_t,
\end{equation}
the last equality follows from (\ref{b1}). Put
\begin{equation} \label{b9}
\bW_x = \diag(\bW_{x,1}, \ldots, \bW_{x,q}).
\end{equation}
Then $\bW_{x,j}$ is a $p_j \times p_j$ positive definite matrix, and
the eigenvalues of $\bW_{x,j}$ are also the eigenvalues of $\bW_x$.
Suppose that $\bW_{x,i}$ and $\bW_{x,j}$ do not share the same
eigenvalues for any $i \neq j$. Then if we line up the
eigenvalues of $\bW_x $ (i.e. the eigenvalues of $\bW_{x,1},
 \ldots, \bW_{x,q}$ combining together)  in the main
diagonal of $\bD$ according to the order of the blocks in $\bW_x$,
$\bGamma_x$ must be a block-diagonal orthogonal matrix of the same
shape as $\bW_x$; see Proposition~\ref{pn:1}(i).
However the order of the eigenvalues is latent, and
any $\bGamma_x$ defined by (\ref{b5}) is nevertheless a
column-permutation of such a block-diagonal orthogonal matrix; see
Proposition~\ref{pn:1}(ii).
Hence each component of $\bGamma_x^\T
\bx_t$ is a linear transformation of the elements in one  of the $q$
subseries only, i.e. the $p$ components of $ \bGamma_y^\T \by_t =
\bGamma_x^\T \bx_t $ can be partitioned
into the $q$ groups such that there exist neither contemporaneous nor
serial correlations across different groups. Thus $\bGamma_y^\T \by_t$ can
be regarded as a permutation of $\bx_t$, and $\bGamma_y$ can be viewed as
a column-permutation of $\bA$; see the discussion below (\ref{b8}). This
leads to the following two-step estimation for $\bA$ and $\bx_t$:

\begin{quote}
\begin{description}
\item[Step 1.] Let $\wh \bS$ be an estimator for $\bW_y$. Calculate a $p\times p$ orthogonal matrix
$\wh \bGamma_y$ with the
columns being the orthonormal eigenvectors of $\wh \bS$.

\item[Step 2.] The columns of $\wh \bA   = (\wh \bA_1,
\ldots, \wh \bA_q)$  are a permutation of  the columns of $\wh
\bGamma_y$ such that $\wh \bx_t = \wh \bA^\T \by_t $ is segmented
into $q$ uncorrelated subseries $\wh \bx_t^{(j)} = \wh
\bA_j^\T\by_t$, $j=1, \ldots, q$.
\end{description}
\end{quote}
Step 1 is the key, as it provides an estimator for $\bA$ except that
the columns of the estimator are not grouped together according to
the latent segmentation. The estimator $\wh \bS$ will be
discussed in Section \ref{sec3}. The permutation in Step 2 above can
be carried out in principle by visual observation: plot cross
correlogram of $\wh \bz_t \equiv \widehat{\bGamma}^\T_y\by_t$
(using, for example, {\sl R}-function {\tt acf}); see
Fig \ref{7cityTemACF}(b).
We then put those components of
$\wh \bz_t$ together when there exist significant cross-correlations
(at any lags) between those component series. Then $\wh \bA$ is
obtained by re-arranging the order of the columns of $\wh \bGamma_y$
accordingly.

\begin{remark}
(i) Appropriate precaution should be exercised in the visual observation
stated above. First the visual observation become impractical when $p$ is
large. Furthermore most correlogram plots produced by statistical
packages (including {\sl R})  use the confidence bounds at $\pm
1.96/\sqrt{n}$ for sample cross-correlations of two time series.
Unfortunately those bounds are only valid if at least one of the two
series is  white noise. In general, the confidence bounds depend on the
autocorrelations of the two series. See Theorem 7.3.1 of
\cite{BrockwellDavis_1996}. In Section~\ref{sec22}, we will describe how
the permutation can be performed without the benefit of
visual observation for the cross correlogram of $\wh \bz_t$.
\cite{LedoitWolf_2004} and \cite{PaparoditisPolitis_2012} provide more
modern approaches to view correlations.

(ii) $\bW_y$ defined in (\ref{Wy}) combines the information over different time lags together. In practice we need to specify the integer $k_0$. Note that all terms on the right-hand side of (\ref{Wy}) is non-negative definite. Hence there is no information cancellation over different lags. This makes the method insensitive to the choice of $k_0$. In practice a small $k_0$ is often sufficient, as long as the first $k_0$ lags carry sufficient information on the latent block diagonal structure even when the auto/cross-correlations beyond lag $k_0$ are still significant.   The examples in Section~\ref{sec5} lend further support to this assertion.

\end{remark}

\begin{proposition}\label{pn:1}
{\rm(i)} The orthogonal matrix $\bGamma_x$ in {\rm(\ref{b5})} can be taken as a block-diagonal orthogonal matrix with the same block structure as $\bW_x$.

{\rm(ii)} An orthogonal matrix $\bGamma_x$ satisfies {\rm(\ref{b5})} if
and only if
its columns are a permutation of the columns of a block-diagonal
orthogonal matrix described in {\rm(i)},
provided that any two different blocks $\bW_{x,i}$ and
$\bW_{x,j}$ do not share the same eigenvalues.
\end{proposition}

Proposition 1(ii) requires that the $q$ blocks of
$\bW_x$ do not share the same eigenvalue(s). However it does not
rule out the possibility that each block $\bW_{x,j}$ may have multiple eigenvalues.
 When different blocks share the same eigenvalue(s),
Proposition~\ref{pn:1} still holds with $\bW_x$ replaced by $\bW_x^\star$ which is also a block diagonal matrix with fewer than $q$ blocks obtained by combining together those $\bW_{x,j}$'s sharing at least one common eigenvalue into one larger block. This means that the proposed method will not be able to separate, for example,
 $\bx_t^{(1)}$ and $\bx_t^{(2)}$ if $\bW_{x,1}$ and $\bW_{x,2}$ share at least
one common eigenvalue.

\subsection{Permutation} \label{sec22}

\subsubsection{Permutation rule} \label{sec220}
The columns of $\wh \bA$ is a permutation of the columns of $\wh
\bGamma_y$. The permutation is determined by grouping the components
of $\wh \bz_t = \wh \bGamma_y^\T\by_t$ into $q$ groups, where $q$ and the cardinal numbers
of those groups are unknown. Write $\wh \bz_t=(\wh z_{1,t},\ldots,\wh z_{p,t})^\T$. Let $\rho_{i,j}(h)$ denote the cross correlation between the two component series $\widehat{z}_{i,t}$ and $\wh z_{j,t}$ at lag $h$.
We say $\wh z_{i,t}$ and $\wh z_{j,t}$ {\sl connected} if the
multiple null hypothesis
\begin{equation} \label{b10}
H_0: \rho_{i,j}(h) = 0 \quad \mbox{for any $h =0, \pm 1, \pm 2, \ldots,
\pm m$}
\end{equation}
is rejected, where $m\ge 1$ is a prescribed integer. Thus there exists significant evidence indicating non-zero correlations between two connected component series. Hence those components should be put in the same group. We may take $m=20$, or $m$ sufficiently large but smaller than $n/4$, in the spirit of the rule of thumb proposed by Box and Jenkins (1970, p.30), as we exclude long memory processes in this paper. Note that the autocorrelations of stationary (causal) VARMA processes decay exponentially fast. The permutation in Step 2 in Section~\ref{sec21} can be performed as follows.

\begin{quote}
\begin{enumerate}
\item[i.]
Start with the $p$ groups with each group containing  one component
of $\wh \bz_t$ only.
\item[ii.]
Combine two groups together if one connected pair are split over
the two groups.
\item[iii.] Repeat Step ii above until
all connected pairs are within one group.
\end{enumerate}
\end{quote}
We introduce below two methods for identifying the connected pair
components of $\wh\bz_t =\wh \bGamma_y^\T\by_t$.

\subsubsection{Maximum cross correlation method}
\label{sec221}

One natural way to test hypothesis $H_0$ in (\ref{b10}) is to use the maximum cross correlation over the lags between $-m$ and $m$:
\begin{equation} \label{b19}
\widehat{L}_n(i,j)=\max_{|h|\leq m}|\widehat{\rho}_{i,j}(h)|,
\end{equation}
where $\widehat{\rho}_{i,j}(h)$ is the sample cross correlation between $\widehat{z}_{i,t}$ and $\widehat{z}_{j,t}$ at lag $h$. We would reject $H_0$ for the pair $(\wh z_{i,t}, \wh z_{j,t})$ if
$\widehat{L}_n(i,j)$ is greater than an appropriate threshold value.

Instead of conducting multiple tests for each of the $p_0 \equiv
p(p-1)/2$ pairs components of $\wh \bz_t$, we propose a ratio-based
statistic to single out those pairs for which $H_0$ will be rejected. To
this end, we re-arrange the $p_0$ obtained
$\widehat{L}_n(i,j)$'s in the descending order: $\wh
L_1 \ge \cdots \ge \wh L_{p_0}$. Define
\begin{equation} \label{b14}
\wh r = \arg \max_{1\le j < c_0p_0} \wh L_j /\wh L_{j+1},
\end{equation}
where $c_0 \in (0, 1)$ is a prescribed constant. In all the numerical
examples in Section \ref{sec5} and the supplementary material [\cite{CGY_supple}] we use $c_0 = 0.75$. We reject $H_0$
for the pairs corresponding to $\wh L_1, \ldots,  \wh L_{\wh r}$.

The intuition behind this approach is as follows. Suppose among in
total $p_0$ pairs of the components of $\bx_t$ there are $r$
connected pairs only.  Arrange the true maximum cross
correlations in the descending order: $L_1 \ge \cdots \ge L_{p_0}$.
Then $L_r >0$ and $L_{r+1}=0$, and the ratio $L_j /L_{j+1}$ takes
value $\infty$ for $j=r$. This motivates the estimator $\wh r$
defined  in (\ref{b14}) in which we exclude some minimum $\wh L_j$
in the search for $\wh r$ as $c_0 \in (0, 1)$. This is to avoid the
fluctuations due to the ratios of extreme small values.
This causes little loss in information as, for example,
$0.75p_0$ connected pairs would likely group most, if not all,
component series together; see, e.g., Example~\ref{ex-measles} in
Section \ref{sec5}. The similar idea has been used in
defining the factor dimensions in \cite{LamYao_AOS_2012} and
\cite{ChangGuoYao_2013}.

To state the asymptotic property of the above approach, we use a
graph representation. Let the graph contain $p$ vertexes $\wh V = \{ 1,
\ldots, p\}$, representing $p$ component series of $\wh\bz_t$. Define an
edge connecting vertexes $i$ and $j$ if $H_0$ in (\ref{b10}) for $(\wh z_{i,t}, \wh z_{j,t})$ is rejected by the above
ratio method. Let $\wh E_n$ be
the set consisting all those edges. Let $V =\{ 1, \ldots, p\}$ represent
the $p$ component series of $\bz_t = \bGamma_y^\T \by_t$ defined in
(\ref{b6}), and write $\bz_t=(z_{1,t}, \ldots, z_{p,t})^\T$. Define
\[
E = \Big\{ (i,j): \max_{|h| \le m} |\cor(z_{i,t+h}, z_{j,t})| >0, \,1\leq i<j\leq p \Big\}.
\]
Each $(i,j)\in E$ can be reviewed as an edge. The graph
$(\wh V, \wh E_n)$ is a consistent estimate for the graph $(V, E)$; see
Proposition \ref{prop2} below. To avoid the  technical difficulties
in dealing with `0/0', we modify (\ref{b14}) as follows:
\begin{equation} \label{b15}
\wh r = \arg \max_{1\le j < p_0} (\wh L_j + \de_n)/(\wh L_{j+1} + \de_n),
\end{equation}
where $\de_n >0 $ is a small constant. Assume
\begin{equation*}\label{eq:low} \min_{(i,j)\in E}\max_{|h| \le m}
|\cor(z_{i,t+h}, z_{j,t})|\geq \epsilon_n
\end{equation*}
for some $\epsilon_n>0$ and $n\epsilon_n^2\rightarrow\infty$. Write
\begin{equation} \label{bb16}
\varpi_n = \min_{1\leq i<j\leq
p}\min_{\lambda\in\sigma(\bW_{x,i}),\mu\in\sigma(\bW_{x,j})} |\la - \mu |,
\end{equation}
where $\bW_{x,i}$ is defined in (\ref{b9}), $\sigma(\bW_{x,i})$ denotes
the set consisting of all the eigenvalues of $\bW_{x,i}$.
 Here $\epsilon_n$ denotes the weakest signal to be
identified in $E$, and $\varpi_n$ is the minimum difference between the
eigenvalues from the different diagonal blocks in $\bW_x$.
Arrange the true maximum cross correlations of
$\bz_t$ in the descending order $L_1\geq\cdots\geq L_{p_0}$ and define
\[
\chi_n=\max_{1\leq j<r-1}L_j/L_{j+1},
\]
where $r=|E|$.
 Recall that $\wh\bS$ is
the estimator for $\bW_y$ used in Step 1 in
Section~\ref{sec21}. Let
\begin{equation}\label{eq:sampleautoc} \wh \bSigma_y(h) =
{1\over n} \sum_{t=1}^{n-h} (\by_{t+h} - \bar \by) (\by_t - \bar
\by)^\T~~{\rm and}~~ \bar \by = {1 \over n} \sum_{t=1}^n \by_t.
\end{equation}
Now we state the consistency
in Proposition~\ref{prop2},  which requires
$\varpi_n>0$ [see Proposition~\ref{pn:1}(ii)].
The proof of Proposition~\ref{prop2} is similar to
that of Theorem 2.4 of \cite{ChangGuoYao_2013}, and is therefore omitted.

\begin{proposition} \label{prop2}
Let $\chi_n\delta_n=o(\epsilon_n)$ and
 $\varpi_n^{-1}\|\widehat{\bS}-\bW_y\|_2=o_p(\de_n)$.
Let the singular values of $\wh \bSigma_y(h)$ be uniformly bounded
away from $\infty$ for all $|h|\leq m$. Then for $\wh r$ defined in
{\rm(\ref{b15})}, it holds that $\mathbb{P}(\wh E_n = E ) \to 1$.
\end{proposition}

\begin{remark} (i) The inserting of
 $\de_n$ in the definition of $\wh r$
in (\ref{b15}) is to avoid the undetermined ``0/0" cases.
In practice, we use $\wh r$ defined by
(\ref{b14}) instead, but with the search restricted to
$1 \le j \le c_0 p_0$, as $\de_n$ subscribed in
Proposition~\ref{prop2} is unknown. The simulation results reported in the
supplementary material [\cite{CGY_supple}] indicate that (\ref{b14}) works reasonably well. See also
\cite{LamYao_AOS_2012} and  \cite{ChangGuoYao_2013}.

(ii) The uniform boundedness for the singular values of
$\wh \bSigma_y(h)$ was used to simplify the presentation. If $
\max_{|h|\leq m}\|\widehat{\bSigma}_y(h)\|_2=O_p(\nu_n)
$ for some diverging $\nu_n$, we require
the condition  $\varpi_n^{-1}\nu_n\|\widehat{\bS}-\bW_y\|_2=o_p(\de_n)$.

(iii) The finite sample performance can be improved by prewhitening each
component series $\wh z_{i,t}$ first. Then the
asymptotic variance of $\wh \rho_{i,j}(h) $ is
$1/n$ as long as $\cor(z_{i, t+h}, z_{j,t})=0$, see Corollary 7.3.1 of
\cite{BrockwellDavis_1996}. This makes the maximum cross correlations for
different pairs more comparable. Note that two weakly stationary time
series are correlated if and only if their prewhitened series are correlated.

\end{remark}

\subsubsection{FDR based on multiple tests}
\label{sec222}

Alternatively we can identify the connected pair components of $\wh \bz_t$  by
a false discovery rate (FDR) procedure built
on the multiple tests for cross correlations of each pair series.

In the same spirit of Remark 2(iii), we first prewhiten each component series of $\wh \bz_t$ separately, and then look into the cross correlations of the prewhitened series which are white noise. Thus we only need to test hypothesis (\ref{b10}) for two white noise series.

To fix the idea, let $\xi_t$ and $\eta_t$ denote two white noise series. Let $\rho(h) = \cor( \xi_{t+h}, \eta_t)$ and
$
\wh \rho(h)$
be its sample analogue. By Theorem 1  of
\cite{BrockwellDavis_1996},
$\wh \rho(h_1)$  and  $\wh \rho(h_2)$, for any $h_1\ne h_2$,
are asymptotically independent as $n \to \infty$, provided that
$\rho(h) = 0 $ for all $h$,  and the underlying processes are
Gaussian. Hence the $P$-value for testing a simple null hypothesis
$\rho(h) =0$ based on statistic $\wh \rho(h)$ is approximately equal
to
$
p_h = 2 \Phi( - \sqrt{n} |\wh \rho(h)|),
$
where $\Phi(\cdot)$ denotes the distribution function of $N(0, 1)$.
Let $p_{(1)} \le \cdots \le p_{(2m+1)}$ be the order statistics of
$\{p_h:  h =0, \pm 1, \ldots, \pm m\}$. As these $P$-values are
approximately independent for large $n$, a multiple test at the
significant level $\alpha \in (0, 1)$ rejects $H_0$, defined in (\ref{b10}), if
$
p_{(j)} \le j \alpha/ (2m +1)$ for at least one
$ 1 \le j \le 2m +1.
$
See \cite{Simes_1986} for details. \cite{SarkarChang_1997}
showed that it is still a valid test at the level $\alpha$ if $\wh
\rho(h)$, for different $h$, are positive-dependent.
Hence {\sl the $P$-value for this multiple test} for the null
hypothesis $H_0$ is
$
P = \min_{1\le j \le 2m+1} p_{(j)}\, (2m+1)/j.
$
The prewhitening is necessary in conducting the multiple test above, as otherwise $\wh \rho(h_1)$ and $ \wh \rho(h_2)$ ($h_1\ne h_2$) are not asymptotically independent.

We can
calculate the $P$-value  for testing $H_0$ in (\ref{b10}) for each pair of the
components of $\wh \bz_t$, resulting in the total $p_0 \equiv
p(p-1)/2$ $P$-values. Arranging those $P$-values
 in
ascending order: $P_{(1)} \le
 \cdots \le P_{(p_0)}$. Let
\begin{equation} \label{b13}
d = \max\{ k: 1\le k \le p_0, \; P_{(k)} \le k \beta/p_0 \}
\end{equation}
for a given small $\beta \in (0, 1)$. Then the FDR procedure with the error rate controlled under
 $\beta$ rejects the hypothesis $H_0$ for the $d$ pairs of the components of $\wh \bz_t$
corresponding to the $P$-values $P_{(1)}, \ldots, P_{(d)}$, i.e. those
$d$ pairs of components are connected. Since the $P$-values $P_j$'s are
no longer independent, the $\beta$ in (\ref{b13}) no longer admits the standard
FDR interpretation. Nevertheless  the $P$-values $P_{(1)}, \ldots, P_{(d)}$
give another way (in addition to the maximum cross correlation) to rank
the pairs of the components of $\wh \bz_t$
according to the strength of the cross correlations. In fact the ranking of the pairs
in terms of the correlation strength matters most as far
as the dimension-reduction is concerned.
See, e.g., Table\ref{tab1} for Example~\ref{ex-measles} in Section \ref{sec5}.

 \setcounter{equation}{0}

\section{Theoretical properties}
\label{sec3}

To gain more appreciation of the new methodology, we now investigate the
asymptotic properties of the estimator $\wh \bGamma_y$  derived in Step 1
of the proposed method in Section 2.1. More precisely we will show that
there {\sl exists} a permutation transformation which permutes the column
vectors of $\wh \bGamma_y$, and the resulting new orthogonal matrix,
denoted as $\wh \bA = (\wh \bA_1, \ldots, \wh \bA_q)$, is an adequate
estimator for the transformation matrix $\bA$ in (\ref{b1}) in the sense
that $\calM(\wh \bA_j)$ is consistent to $\calM(\bA_j)$ for each
$j=1, \ldots, q$. In this section, we treat this permutation
transformation as an `oracle'. In practice it is identified either by a
visual observation or by the
methods presented in Section~\ref{sec22}. Our goal here is to show
that $\wh \bGamma_y$ is a valid estimator for $\bA$ upto a column permutation.
 We establish the consistency under three
different asymptotic modes:
(i) the dimension $p$ is fixed, (ii) $p= o(n^c)$, and
(iii) $\log p = o(n^c)$, as the sample size $n \to \infty$, where
$c>0$ is a small constant. The convergence rates derived reflect the asymptotic
orders of the estimation errors when $p$ is in different orders
in relation to $n$.

To measure the errors in estimating $\calM(\bA_j)$, we adopt a metric
on the Grassmann manifold of $r$-dimensional
subspaces of $\mathbb{R}^p$: for two $p\times r$ half orthogonal
matrices $\bH_1$ and $\bH_2$ satisfying the condition
$\bH_1^\T\bH_1=\bH_2^\T\bH_2=\bI_r$, the distance between
$\mathcal{M}(\bH_1)$ and $\mathcal{M}(\bH_2)$ is defined as
\begin{equation*} \label{discM}
D(\mathcal{M}(\bH_1),\mathcal{M}(\bH_2))=\sqrt{1-{r}^{-1}\textrm{tr}
(\bH_1\bH_1^\T\bH_2\bH_2^\T)}.
\end{equation*}
Then $D(\mathcal{M}(\bH_1),\mathcal{M}(\bH_2)) \in [0,
1]$. It is equal
to 0 if and only if $\mathcal{M}(\bH_1)=\mathcal{M}(\bH_2)$, and to 1 if and only if
$\mathcal{M}(\bH_1)$ and $\mathcal{M}(\bH_2)$ are orthogonal.
See, for example, \cite{StewartSun_1990} and
\cite{PanYao_2008}.

We always  assume that the weakly stationary process $\by_t$ is $\alpha$-mixing,
i.e. its mixing coefficients $\alpha_{k,p} \to 0$ as $k \to \infty$, where
\begin{equation}\label{eq:a}
 \alpha_{k,p} = \sup_i  \sup_{ A \in \mathcal {F}_{-\infty}^i, \; B \in
\mathcal {F}_{i+k}^\infty} | \mathbb{P}(A \cap B) - \mathbb{P}(A) \mathbb{P}(B) |,
\end{equation}
and $\mathcal {F}_i^j$ is the $\sigma$-field generated by $\{{\bf y}_t:
\; i \leq t \leq j \}$. In sequel, we denote by
$\sigma_{i,j}^{(k)}$ the $(i,j)$-th element of $\bSigma_y(k)$ for each
$i,j=1,\ldots,p$ and $k=1,\ldots,k_0$.
The $\alpha$-mixing is a mild condition on `asymptotic independence'. It rules out, for
example, long memory processes. On the other hand, many time series including
causal ARMA processes with continuously distributed
innovations are $\alpha$-mixing with exponentially decaying mixing coefficients. See, e.g.
Section 2.6.1 of \cite{FanYao_2003} and the references within. Let $\bmu\equiv \mathbb{E}(\by_t)$. Write $\by_t=(y_{1,t},\ldots,y_{p,t})^\T$ and $\bmu=(\mu_1,\ldots,\mu_p)^\T$.

\subsection{Asymptotics when $n \to \infty$ and $p$ fixed}\label{se:fixed}

When the dimension $p$ is fixed, we estimate $\bW_y$ defined in
(\ref{Wy}) by the plug-in estimator
\begin{equation} \label{eq:hWy}
\wh \bS =  \bI_p +\sum_{k=1}^{k_0} \wh \bSigma_y(k) \wh
\bSigma_y(k)^\T,
\end{equation}
where $\wh \bSigma_y(k)$ is defined in (\ref{eq:sampleautoc}).
We show that the standard $\sqrt{n}$ convergence rate prevails as now $p$ is fixed.
We introduce some regularity conditions first.

\begin{as}\label{as:fm}
It holds that $\sup_t\max_{1\leq i\leq p}\mathbb{E}(|y_{i,t}-\mu_i|^{2\gamma})\leq K_1$ for some constants $\gamma>2$ and $K_1>0$.
\end{as}

\begin{as}\label{as:falpha}
The mixing coefficients $\alpha_{k,p}$ defined in (\ref{eq:a})
satisfy the condition
% $\sum_{k=1}^\infty\alpha_k^{1-2/\gamma}<\infty$,
 $ \sum_{k=1}^\infty\alpha_{k,p}^{1-2/\gamma}
<\infty$,
where $\gamma>2$ is given in Condition \ref{as:fm}.
\end{as}

\begin{theorem}\label{tm:f}
Under Conditions {\rm\ref{as:fm}} and {\rm\ref{as:falpha}}, if $\varpi_n>0$ with fixed $p$
in {\rm(\ref{bb16})}, then
$
\max_{1\le j \le q}D(\mathcal{M}(\widehat{{\bf
A}}_j),\mathcal{M}({\bf A}_j))=O_p(n^{-1/2}),
$
where the columns of $\wh \bA = (\wh \bA_1, \ldots, \wh \bA_q)$ are
a permutation of the columns of $\wh \bGamma_y$ obtained by $\widehat{\bS}$ defined in {\rm(\ref{eq:hWy})}.
\end{theorem}

\begin{remark}
This result can be extended to non-stationary case. For $p$-dimensional non-stationary time series $\by_t$, we assume that $\by_t=\bA\bx_t$ where $\bx_t$ satisfies (\ref{b7}). Let
$
\bSigma_y(k)=(n-k)^{-1}\sum_{t=1}^{n-k}\textrm{Cov}(\by_{t+k},\by_t)$ and $\bSigma_x(k)=(n-k)^{-1}\sum_{t=1}^{n-k}\textrm{Cov}(\bx_{t+k},\bx_t)
$, which can be viewed as the extension of the
conventional autocovariance for stationary process to non-stationary
case. Then (\ref{b3}) still holds. Following the same arguments stated in
\cite{ChangGuoYao_2013}, it can be shown that there exists $\wh \bA =
(\wh \bA_1, \ldots, \wh \bA_q)$ such that Theorem 1 holds, where the
columns of $\wh \bA$ is a permutation of the columns of $\wh \bGamma_y$,
and the columns of $\wh \bGamma_y$ are the orthonormal eigenvectors of
$\widehat{\bS}$ defined in (\ref{eq:hWy}) with $\widehat{\bSigma}_y(k)$
specified in (\ref{eq:sampleautoc}).
\end{remark}

\subsection{Asymptotics when $n\to \infty$ and $p= o(n^c)$}
\label{sec32}

In the contemporary statistics dealing with large data, conventional wisdom
assumes that $p$ diverges together
with $n$. Since $\|\wh
\bS-\bW_y\|_F=O_p(pn^{-1/2})$ for $\widehat{\bS}$ defined in (\ref{eq:hWy}), it is necessary that $p=o(n^{1/2})$
in order to retain the consistency (but with a slower convergence rate
than root-$n$). This means that $p$ can only be
as large as $p=o(n^{1/2})$ if we do not entertain any additional
assumptions on the underlying structure. In order to deal with
large $p$, we impose in Condition~\ref{as:sparse} below the sparsity on
the transformation matrix~$\bA$ in (\ref{b1}).

\begin{as}\label{as:sparse}
Write $\bA=(a_{i,j})$.
It holds that
$
\max_{1\le j \le p}\sum_{i=1}^p|a_{i,j}|^{\iota}\leq s_1$ and $\max_{1\le i \le p}\sum_{j=1}^p|a_{i,j}|^{\iota}\leq s_2,
$ for some constant $\iota\in[0,1)$,
where $s_1$ and $s_2$ may diverge together with $p$.
\end{as}

When $p$ is fixed, Condition 3 holds for $s_1=s_2=p$ and any $\iota\in[0,1)$, as $\bA$ is an orthogonal matrix. For large $p$,  $s_1$ and $s_2$ control the degree of the sparsity of the columns and the rows of $\bA$ respectively.
 A small $s_1$ entails
that each component series of ${\bf x}_t$ only
contributes to a small fraction of the components of ${\bf y}_t$. A
small $s_2$ entails  that each component of  ${\bf y}_t$ is a linear
combination of a small number of the components of ${\bf x}_t$. The
sparsity of $\bA$ is also controlled by constant $\iota$: the
smaller $\iota$ is, the more sparse $\bA$ is. We will show that the
stronger sparsity leads to the faster convergence for our estimator;
see Remark 4(ii) below.

If $p$ diverges faster than $n^{1/2}$, the sample autocovariance
matrix $\wh \bSigma_y(k) =
(\wh\sigma_{i,j}^{(k)})$, given in
(\ref{eq:sampleautoc}), is no longer a consistent estimator for
$\bSigma_y(k)$. Inheriting the spirit of threshold estimator for
large covariance matrix by
\cite{BickelLevina_2008}, we employ the following threshold
estimator instead:
\begin{equation}\label{eq:threshold}
T_u(\wh\bSigma_y(k))=\big(\wh \sigma_{i,j}^{(k)}\mathbb{I}\{|\wh
\sigma_{i,j}^{(k)}|\geq u\}\big),
\end{equation}
where $\mathbb{I}(\cdot)$ is the indicator function, $u= M\max\{p^{2/l}n^{-(l-1)/l},(n^{-1}\log p)^{1/2}\}$ is the threshold level, and $M>0$ is a constant. The threshold value is due to the fact $\max_{1\leq i,j\leq p}|\wh\sigma_{i,j}^{(k)}-\sigma_{i,j}^{(k)}|=O_p(\max\{p^{2/l}n^{-(l-1)/l},(n^{-1}\log p)^{1/2}\})$, see Lemma 4 in the supplementary material [\cite{CGY_supple}]. Consequently, we define now
\begin{equation} \label{Wthre}
\wh \bS \equiv
\wh\bW_{y}^{(\textrm{thre})}=
\bI_p+\sum_{k=1}^{k_0}T_u(\wh\bSigma_y(k))T_u(\wh\bSigma_y(k))^\T.
\end{equation}
Lemma~7 in the supplementary material [\cite{CGY_supple}] shows that $\wh\bW_{y}^{(\textrm{thre})}$ is
a consistent estimator for $\bW_{y}$, which requires a  stronger version
of Conditions 1 and 2 as now $p$ diverges together with $n$.

\begin{as}\label{as:ptail}
As $x\rightarrow\infty$, it holds that
$
\sup_t\max_{1\le i \le p}
\mathbb{P}(|y_{i,t}-\mu_i|>x)=O\{x^{-2(l+\tau)}\}$ for some constants $l>2$ and $\tau>0$.
\end{as}

\begin{as}\label{as:palpha}
The mixing coefficients $\alpha_{k,p}$ given in (\ref{eq:a}) satisfy the condition
$
\sup_{p\geq1}\alpha_{k,p}=O\{k^{-(l-1)(l+\tau)/\tau}\}$ as $k\rightarrow\infty$,
where $l$ and $\tau$ are given in Condition \ref{as:ptail}.
\end{as}

Conditions \ref{as:ptail} and \ref{as:palpha} ensure the Fuk-Nagaev type inequalities for $\alpha$-mixing processes, see \cite{Rio_2000} and \cite{LiuXiaoWu_2013}. For $j=1, \ldots, q$, define
\begin{equation} \label{rhoj}
\rho_j=\min_{i\ne j } \min_{\lambda\in\sigma(\bW_{x,i}), \mu\in\sigma(\bW_{x,j})}|\lambda-\mu|.
\end{equation}
Put
\begin{equation}\label{eq:delta}
\delta=s_1s_2\max_{1\leq j\leq q}p_j~~\textrm{and}~~\kappa=\max_{1\leq k\leq k_0}\|\bSigma_x(k)\|_2.
\end{equation}
Now we let $\wh \bS = \wh\bW_y^{(\thre)}$ in Step 1 in our estimation method. Then we have the following theorem.

\begin{theorem}\label{tm:p}
Define $\vartheta_n=\max\{p^{2/l}n^{-(l-1)/l},(n^{-1}\log p)^{1/2}\}$ with $l$ given in Condition {\rm\ref{as:ptail}}. Under Conditions {\rm\ref{as:sparse}}, {\rm\ref{as:ptail}} and {\rm\ref{as:palpha}}, if $\min_{1\leq j\leq q}\rho_j>0$ for $\rho_j$ defined in {\rm(\ref{rhoj})},
 and $p=o\{n^{(l-1)/2}\}$, then there exists an $\wh \bA = (\wh \bA_1, \ldots, \wh \bA_q)$ of which the columns are a permutation of the columns of $\wh \bGamma_y$ obtained by $\widehat{\bS}$ defined in {\rm(\ref{Wthre})} with the threshold level $u\asymp \vartheta_n$ in {\rm(\ref{eq:threshold})}, such that
$
\max_{1\le j \le q}\rho_jD(\mathcal{M}(\widehat{{\bf A}}_j),\mathcal{M}({\bf A}_j))= O_p\{\kappa\vartheta_n^{1-\iota}\delta+\vartheta_n^{2(1-\iota)}\delta^2\}.
$
\end{theorem}

\begin{remark} (i) Theorem \ref{tm:p} presents the uniform convergence rate for $\rho_jD(\mathcal{M}(\widehat{{\bf A}}_j),\mathcal{M}({\bf A}_j))$. As $\rho_j$ measures the minimum difference between the eigenvalues of $\bW_{x,j}$ and those of the other blocks, it is intuitively clear that the smaller this difference is, more difficult the estimation for $\calM(\bA_j)$ is.

(ii) As $\bSigma_y(k) = \bA \bSigma_x(k) \bA^\T$, the largest block size
$S_{\max}=\max_{1\leq j\leq q}p_j$ and the sparsity of $\bA$ determine
the sparsity of $\bSigma_y(k)$. Lemma 6 in supplementary material shows that
the sparsity of $\bSigma_y(k)$ can be evaluated by $\delta$ defined in
(\ref{eq:delta}). A small value of $S_{\max}$ represents a high degree of
sparsity for $\bSigma_x(k)$ and, thus, also for $\bSigma_y(k)$,
while the sparsity of $\bA$ is reflected by $\iota$, $s_1$ and $s_2$; see
Condition 3 and the comments immediately below it.
 The convergence rates specified in Theorem~\ref{tm:p} contain
factors $\delta$ or $\delta^2$. Hence the more sparse
$\bSigma_y(k)$ is (i.e. the smaller $\delta$ is), the faster the
convergence is.

(iii) With the sparsity imposed in Condition~\ref{as:sparse}, the dimension of time series can be as large as $p=o\{n^{(l-1)/2}\}$, where $l>2$ is determined by the tail probabilities described in Condition 4.

(iv) Similar to Theorem 1, the result  in Theorem 2 can also be extended to non-stationary case. See
Remark 3.

(v) As discussed in Remark 4(ii), the factor $\delta$ reflects the sparsity of $\bSigma_y(k)$ for each $k=1,\ldots,k_0$. See Lemma 5 in the supplementary material [\cite{CGY_supple}] for details. Instead of requiring Condition \ref{as:sparse}, if we impose the sparsity condition on each $\bSigma_y(k)$ such that $
\max_{1\leq j\leq p}\sum_{i=1}^p|\sigma_{i,j}^{(k)}|^\iota\leq
s_3$ and $\max_{1\leq i\leq p}\sum_{j=1}^p|\sigma_{i,j}^{(k)}|^\iota\leq
s_3$
for some $\iota\in[0,1)$, the convergence rate specified in Theorem \ref{tm:p} changes to $O_p\{\kappa\vartheta_n^{1-\iota}s_3+\vartheta_n^{2(1-\iota)}s_3^2\}$. Under the ideal case $\kappa=O(1)$, $\min_{1\leq j\leq q}\rho_j\asymp q^{-1}$ and $s_3\asymp p^{\zeta}$ for some $\zeta\in[0,1)$, we have $\max_{1\leq j\leq q}D(\mathcal{M}(\widehat{\bA}_j),\mathcal{M}(\bA_j))=O_p(p^{\zeta}q\vartheta_n^{1-\iota})$ provided that $p^{\zeta}\vartheta_n^{1-\iota}=O(1)$. Therefore, if $p^\zeta q\vartheta_n^{1-\iota}=o(1)$, we can estimate each subspace $\mathcal{M}(\bA_j)$ consistently.
\end{remark}

\subsection{Asymptotics when $n \to \infty$ and $\log p = o(n^c)$}\label{se:ultrahigh}

To handle the ultra high-dimensional cases where
$p$ grows at an  exponential rate of $n$, we
need following stronger conditions (than Conditions \ref{as:ptail} and \ref{as:palpha})
 on the decays of the tail probabilities
of $\by_t$ and the mixing coefficients $\alpha_{k,p}$ defined in (\ref{eq:a}). %See below.

\begin{as}\label{as:exptail}
It holds for any $x>0$ and $\|{\bf v}\|_2=1$ that
$
\sup_t \mathbb{P}\{|{\bf v}^\T({\bf y}_t-\bmu)|>x\}\leq K_2\exp(-K_3x^{r_1}),
$
where $K_2, K_3 >0$, and $r_1\in(0,2]$ are constants.
\end{as}

\begin{as}\label{as:expalpha}
It holds for all $k\ge 1$ that $
\sup_{p\geq1}\alpha_{k,p}\leq \exp(-K_4k^{r_2})
$, where
$K_4>0$ and $r_2\in(0,1]$ are some constants.
\end{as}

Condition \ref{as:exptail} requires the tail probabilities
of linear combinations of ${\bf
y}_t$ decay exponentially fast. When $r_1=2$, ${\bf y}_t$ is
sub-Gaussian. It is also intuitively clear that the large $r_1$ and/or $r_2$
would only make Conditions \ref{as:exptail} and/or \ref{as:expalpha}
stronger. The restrictions $r_1 \le 2$ and $ r_2 \le 1$  are
introduced only for the presentation convenience, as
Theorem~\ref{tm:e} below applies to the ultra high-dimensional cases with
\begin{equation} \label{gamma1}
\log p = o\{n^{\varrho/(2- \varrho)}\}, \quad {\rm where} \;\; \varrho = 1/(2 r_1^{-1} + r_2^{-1}).
\end{equation}

We still use $\wh \bS=\wh\bW_y^{(\thre)}$ defined in (\ref{Wthre})
in Step 1 of our procedure. But now
the threshold value is set at $u=M(n^{-1}\log p)^{1/2}$ in
(\ref{eq:threshold}),
 as Lemma 8 in the supplementary material [\cite{CGY_supple}] indicates that $\max_{1\leq i,j\leq p}|\widehat{\sigma}_{i,j}^{(k)}-\sigma_{i,j}^{(k)}|=O_p\{(n^{-1}\log
p)^{1/2}\}$ when $p$ is specified by (\ref{gamma1}). Recall that $\delta$ and $\kappa$ are defined in (\ref{eq:delta}). Now we are ready to state the asymptotic results.

\begin{theorem}\label{tm:e}
Under Conditions {\rm\ref{as:sparse}}, {\rm\ref{as:exptail}} and {\rm\ref{as:expalpha}}, if $\min_{1\leq j\leq q}\rho_j>0$ for $\rho_j$ defined in {\rm(\ref{rhoj})} and $p$ specifies {\rm(\ref{gamma1})}, then there exists an $\wh \bA = (\wh \bA_1, \ldots, \wh \bA_q)$ of which the columns are a permutation of the columns of $\wh \bGamma_y$ obtained by $\widehat{\bS}$ defined in {\rm(\ref{Wthre})} with the threshold level $u\asymp (n^{-1}\log p)^{1/2}$ in {\rm(\ref{eq:threshold})}, such that
$
\max_{1\le j \le q}\rho_jD(\mathcal{M}(\widehat{{\bf A}}_j),\mathcal{M}({\bf A}_j))=O_p\{\kappa(n^{-1}\log p)^{(1-\iota)/2}\delta+(n^{-1}\log p)^{1-\iota}\delta^2\}$.
\end{theorem}

 \setcounter{equation}{0}

\section{Numerical Properties}
\label{sec5}

 The segmentation is only possible if such a latent structure
exists, as assumed in (\ref{b1}) and (\ref{b7}). Two  questions arise
immediately: (i) Is such an
assumption of practical relevance? (ii) What would the proposed method lead to if
the segmentation assumption does not hold? To answer these questions,
we apply the proposed method to four real data sets arising from
different fields.
 We also consider some simulation studies to
illustrate the finite sample properties of the proposed method. Due to the pages limitation, we only present the real data analysis here and report the simulation studies in the supplementary material [\cite{CGY_supple}].

We always standardize the data using the sample covariance matrix, i.e. to replace $\by_t$ by $\{\wh \bSigma_y(0)\}^{-1/2} \by_t$; see (\ref{b2})
 and (\ref{eq:sampleautoc}).
Then the segmentation transformation is $\wh\bx_t = \wh\bB \by_t$, where
$\wh\bB = \widehat{\bGamma}_y^\T \{\wh \bSigma_y(0)\} ^{-1/2}$, and $\wh \bGamma_y$
is the $p\times p$ orthogonal matrix specified in Step 1 in
Section~\ref{sec21} based on the new time series $\{\wh \bSigma_y(0)\}^{-1/2} \by_t$. We always prewhiten each transformed component
series of $\widehat{\bx}_t$ before applying the permutation methods described in
Section~\ref{sec22}. The prewhitening is carried out by fitting each
series an AR model with the order between 0 and 5 determined by AIC. The
resulting residual series is taken as a prewhitened series. We set the
upper bound for the AR-order at 5 to avoid over-whitening with finite
samples.  We always set $c_0=0.75$ in (\ref{b14}) and $k_0=5$ in
computing $\widehat{\bS}$ unless stated explicitly. See Remark 1(ii).

To show the advantages of the proposed TS-PCA transformation, we also
conduct post-sample forecasting and compare the forecasts based
on the original data directly and those via TS-PCA transformation. To
ensure that the comparison is fair and objective, we adopt VAR models with the
order determined by AIC for both the original and the transformed data,
involving no fine-tuning on the form of model and the order
determination, which are inevitably less objective. Note that there is no
universally accepted optimal model for a real data set. We use the {\sl
R}-function {\tt VAR} in the {\sl R}-package {\tt vars} to fit VAR
models. We also report the results from the restricted VAR model (RVAR)
obtained by setting insignificant coefficients to 0 in a fitted VAR
model, using the {\sl R}-function {\tt restrict} in the {\sl R}-package {\tt vars}.

Some useful tips from the real data analysis below are worth
mentioning. First, the segmentation assumption is reasonable for Examples 1, \ref{ex-UNindustry}
and \ref{ex-clothing}. Secondly, when the segmentation assumption is
invalid (Example \ref{ex-measles}), the TS-PCA
transformation leads to approximate segmentations which also improve
the forecasting performance.
Thirdly, when $p$ is large or moderately large it is necessary to
apply appropriate dimension-reduction techniques (such as the
proposed TS-PCA) in order to make use of the dependence
across different series. Finally, the forecasting via the TS-PCA
transformation always outperform that directly based on the original
data in all the real data examples. The reason for this is explained at the end of Section 6.

\vspace{10pt}

 {\sc Example 1}. ({\it Continue}) We continue the analysis with
the monthly temperature data in the 7 cities in China. The  result
reported in Section~\ref{sec21} was obtained with $k_0=5$ in
(\ref{Wy}).  The profile of the segmentation is unchanged for $1 \le
k_0 \le 36$. For $p=7$, we do not need to apply the methods in
Section~\ref{sec22} for permuting the transformed series.
Nevertheless exactly the same grouping is obtained by the
permutation based on the maximum cross correlation method with $1\le
m \le 30$ in (\ref{b10}), or by the
permutation based on FDR with $1\le m \le 30$ and $0.001\% \le \beta
\le 1\%$ in (\ref{b13}).

Forecasting the original time series $\by_t$ can be carried out in
two steps: First we forecast the components of $\wh \bx_t$ using 5
models according to the segmentation, i.e. one VAR for the first
three components, and a univariate AR model for each of the last four
components. Then the forecasted values for $\by_t$ are obtained via
the transformation $ \wh\by_{t} = \wh \bB^{-1} \wh\bx_{t}$. For each of the
last 24 observations in this data set (i.e. the monthly temperatures in
1997 and 1998), we use the data up to the previous month to fit three
forecasting models: the model based on the segmentation (which
is a collection of 5 VAR/AR models for the 5 segmented subseries of $\wh
\bx_t$), the VAR and RVAR models for the original data.
 We difference the original data at lag 12 before fitting
them directly with VAR and RVAR models, to remove the seasonal
components. For fitting the segmented series $\widehat{\bx}_t$, we only difference its
first two component series also at lag 12 since only they have seasonal components. The one-step-ahead forecasts can be
obtained directly from the fitted models. The two-step-ahead
forecasts are obtained based on the plug-in method, i.e. using the
one-step-ahead forecasted values as true values.

For each component series of $\by_t$, we calculate the mean squared errors (MSE)
$
d^{-1} \sum_{h=1}^d (\wh y_{i,n_0+h} - y_{i,n_0+h} )^2
$
for both one-step-ahead and two-step-ahead forecasting, where $\wh y_{i,n_0+h}$ denotes the associated forecast for $y_{i,n_0+h}$ (for this example, $d=24$ and $n_0=n-24$).
 The mean and standard
deviations of those MSEs over the 7 cities are listed in
Table\ref{table7cities}.
 Both the mean and standard deviation of the MSEs
based on TS-PCA are much smaller than those based
on the direct VAR or RVAR models for original data.
To evaluate the sensitivity of the segmentation, we also consider the over-segmentation case with the first two components of $\wh
\bx_t$ as a group (since both of them have strong periodicity) and the
other $5$ components as $5$ individual groups (i.e., six groups with
\{1,\,2\}, \{3\}, \{4\}, \{5\}, \{6\}, \{7\}). An incomplete-segmentation case
with 4 groups (\{1,\,2,\,3\}, \{5,\,6\}, \{4\}, \{7\}) are also considered.
 The results in
Table\ref{table7cities} show that, though the predictions for over-
and incomplete-segmentation are worse than the TS-PCA, they still work better than the direct VAR and RVAR models.

\begin{table}[htb]
\begin{center}
\caption{Post-sample forecast performance of different methods. In Examples \ref{ex-temperature}, \ref{ex-UNindustry} and \ref{ex-clothing}, the presented results are the means and standard deviations (in subscripted bracket) of the
MSEs for one-step-ahead and two-step-ahead forecasts. In Example \ref{ex-measles}, the presented results are the means and standard deviations (in subscripted bracket) of the relative MSEs for one-step-ahead and two-step-ahead forecasts.}
\vspace{1ex}
\begin{tabular}{c|c|cc}
   \hline %\hline
Example   &    Method      &     One-step forecast  & Two-step forecast   \\
   \hline
 &\text{VAR}      &   $ 2.470_{(0.416)}$  & $2.559_{(0.385)} $ \\
%  \cline{2-3}
  &\text{RVAR}    &    $2.530_{(0.398)}$ & $ 2.615_{(0.382)}$\\
%\cline{2-3}
Example \ref{ex-temperature}  &\text{Segmentation}     &   $2.221_{(0.339)}$ & $2.203_{(0.323)}$\\
  & \text{Over-segmentation}     &    $2.417_{(0.348)}$ &  $2.419_{(0.326)}$\\
  & \text{Incomplete-segmentation}     &     $2.421_{(0.343)}$ & $2.422_{(0.325)}$\\
 \hline
 &\text{VAR}      &   $0.615_{(0.741)}$  & $1.168_{(1.182)} $ \\
 &\text{RVAR}    &    $0.606_{(0.711)}$ & $1.159_{(1.232)}$\\
Example 2 & \text{Segmentation}      & $0.588_{(0.708)}$ & $1.154_{(1.163)}$\\
 &\text{Over-segmentation}      &  $0.593_{(0.702)}$ & $ 1.158_{(1.154)}$\\
 & \text{Incomplete-segmentation}      &  $ 0.600_{(0.665)}$ &$ 1.140_{(1.165)}$\\
\hline
 &\text{VAR}      & $0.950_{(0.148 )}$  & $0.726_{(0.328)} $ \\
 & \text{RVAR}   &  $0.962_{(0.138)}$ & $0.796_{(0.277 )}$\\
Example 3 & \text{Segmentation}     & $0.884_{(0.180)}$ & $0.708_{(0.377)}$\\
  &  \text{Over-Segmentation}     & $0.919_{(0.130)}$ & $0.884_{(0.219)}$\\
  &    \text{Incomplete-Segmentation}     & $0.873_{(0.176)}$ & $0.694_{(0.377)}$\\
    \hline
 &\text{Univariate AR}                    & $0.208_{(0.551)}$ & $ 0.194_{(0.539)} $ \\
%  \text{univariate model}                    & $0.067_{(0.309)}$ & $ 0.071_{(0.279)} $ \\
 &\text{VAR}                   & $0.295_{(0.806)}$ & $0.301_{(0.855)} $ \\
Example 4 & \text{RVAR}                 & $0.293_{(0.820)}$ & $0.296_{(0.863)}$\\
 & \text{Segmentation}            & $0.153_{(0.134)}$ & $0.163_{(0.124)}$\\
%    \text{Segmentation2}            & $0.061_{(0.276)}$ & $0.068_{(0.261)}$\\
 & \text{Over-segmentation}     & $0.110_{(0.084)}$ & $0.132_{(0.091)}$
  \\
 & \text{Incomplete-segmentation}     & $0.151_{(0.133)}$ & $0.159_{(0.121)}$\\
 \hline
\end{tabular}
\label{table7cities}
\end{center}
\end{table}

\begin{figure}[tbh]
\begin{center}
\subfigure{\includegraphics[scale=0.75]{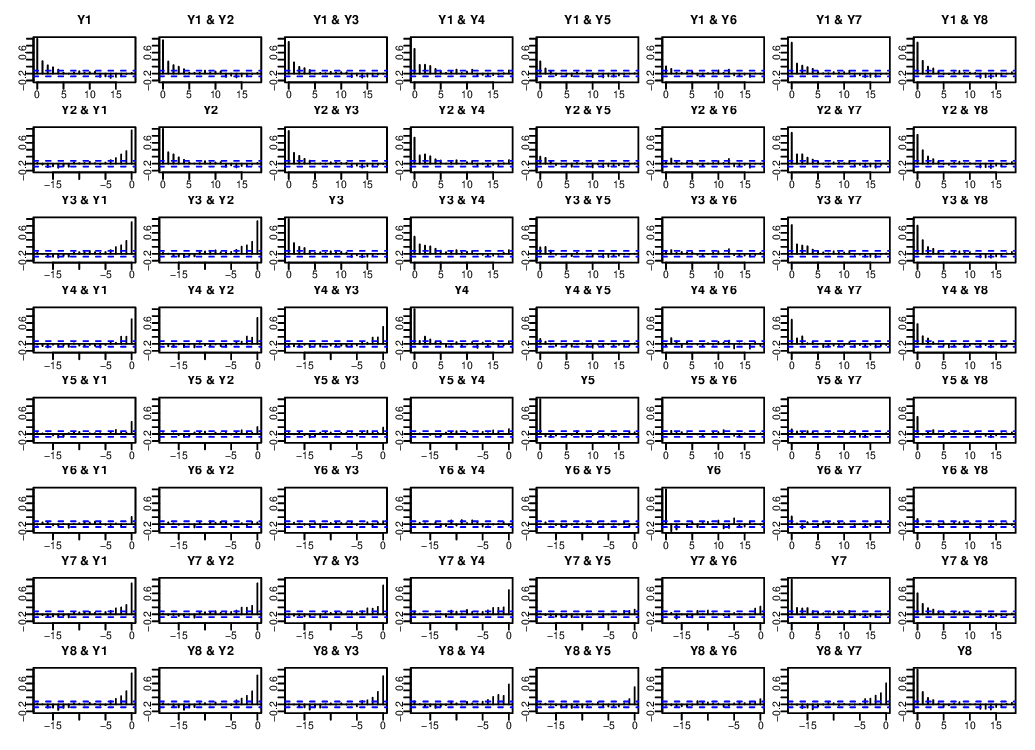}}\\
(a) Cross correlogram of the original $8$ differenced index series.\\
%\subfigure {\includegraphics[scale=0.363]{3exptong1000n5030new}}\\
\subfigure{\includegraphics[scale=0.75]{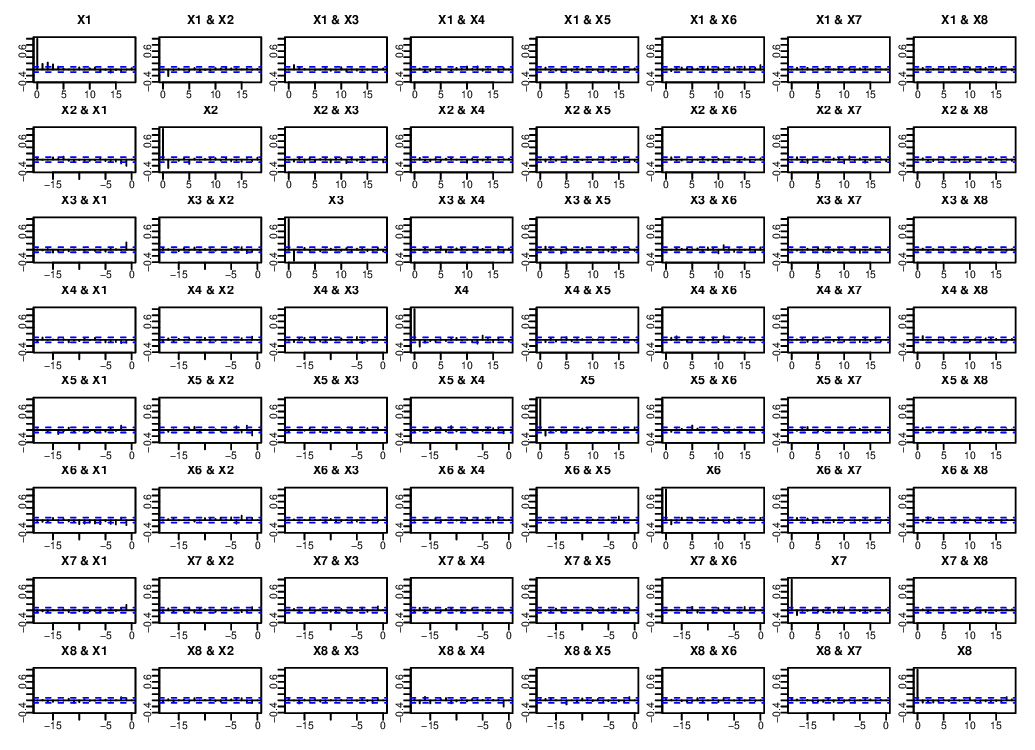}}\\
(b) Cross correlogram of the $8$ transformed component time series.
%\subfigure {\includegraphics[scale=0.363]{3exptong1000n10060}}\\
\setlength{\abovecaptionskip}{0pt}
\end{center}
\caption{Cross correlograms for Example \ref{ex-UNindustry}.}\label{usIndustACF}
\end{figure}

\begin{example} \label{ex-UNindustry}
Now we consider the 8 monthly US Industrial Production  indices in
January 1947 -- December 1993 published by the US Federal Reserve.
The 8 indices concerned are Total Index, Manufacturing
Index, Durable Manufacturing, Nondurable Manufacturing, Mining,
Utilities, Products and Materials.
 Since those index series
exhibit clearly increasing trends, we difference each series first
with their cross correlogram displayed in Fig \ref{usIndustACF}(a).
We apply the TS-PCA to the 8 differenced
indices. The correlogram of the transformed differenced
indices is presented in Fig \ref{usIndustACF}(b).
 A visual observation of
Fig \ref{usIndustACF}(b) would suggest no noticeable cross
correlations in all the panels off the main-diagonal.
But close examination of those off-diagonal panels reveals small but
significant correlations in the panels at the positions (1,\,2),
(1,\,3), (3,\,1) and (8,\,4). This suggests a segmentation with
5 groups:
$\{1,\,2,\,3\}, \{4,\,8\}, \{5\}, \{6\}$ and \{7\}. This segmentation
is also confirmed by the permutation based on FDR with $m=5$ and
$\beta =0.005$. However with $m=5$ and $\beta \in [ 10^{-6}, \,
0.001]$, or $m=20$ and $\beta \in [10^{-6},\,  0.01]$, the
permutation based on FDR leads a segmentation of 7 groups with $\{
1,\,3 \}$ as the only group containing more than one members. The
permutation based on maximum cross correlation method, with $1\le m
\le 20$ in (\ref{b10}), also entails this segmentation of the 7
groups. Looking at the correlogram in Fig \ref{usIndustACF}(b),
there is no need to use large values for $m$. Since those
significant cross correlations are so small, we accept both the
segmentations with the 5 or the 7 groups as viable options for
initial dimension reduction in analysing the original 8-dimensional
time series. We carry out the post-sample forecast comparison in the same manner as in
Example \ref{ex-temperature}. Namely we forecast the monthly indices
in January 1992 -- December 1993 based on the segmentation of
7 groups, direct
VAR and RVAR methods.
  The results are reported in Table\ref{table7cities}.
Similar
to Example \ref{ex-temperature}, we also consider the over-segmentation case with each component as an individual group and the
incomplete-segmentation case with 5 groups (\{1,\,2,\,3\}, \{4,\,8\}, \{5\}, \{6\}, \{7\}). Once again the forecasts via the
segmentation are more accurate than those based on original
data.

\end{example}

\begin{example} \label{ex-measles}
We consider the weekly notified  measles cases in
7 cities in England (i.e. London,  Bristol,
Liverpool,  Manchester,  Newcastle,  Birmingham and  Sheffield) in
1948 -- 1965, before the advent of vaccination. All the 7 series show
biennial cycles,  which is a common feature in measles dynamics
in the pre-vaccination period. This biennial cycling is the
major driving force for the cross correlations among different
component series displayed in Fig \ref{measlesACF}(a). The cross correlogram of the transformed data is displayed in
Fig \ref{measlesACF}(b). Since none of the transformed component series are white noise, the
confidence bounds in
Fig \ref{measlesACF}(b) could be misleading; see Remark 1(i).
\begin{figure}[tbh]
\begin{center}
\subfigure{\includegraphics[scale=0.75]{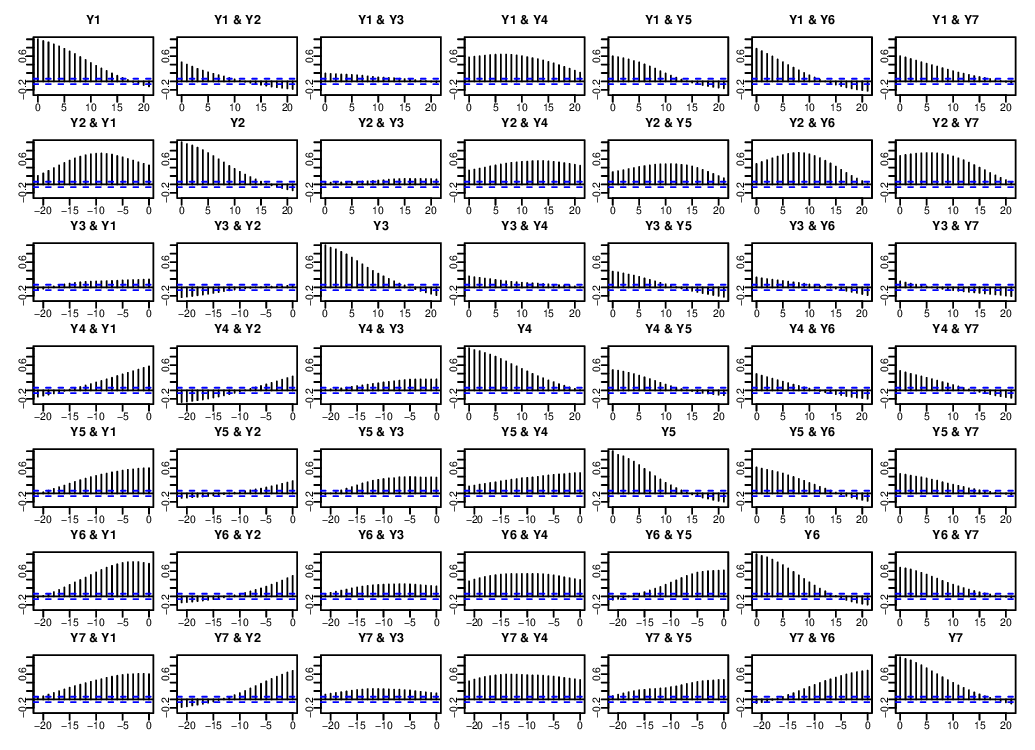}}\\
(a) Cross correlogram of the original $7$ measles series.\\
%\subfigure {\includegraphics[scale=0.363]{3exptong1000n5030new}}\\
\subfigure{\includegraphics[scale=0.75]{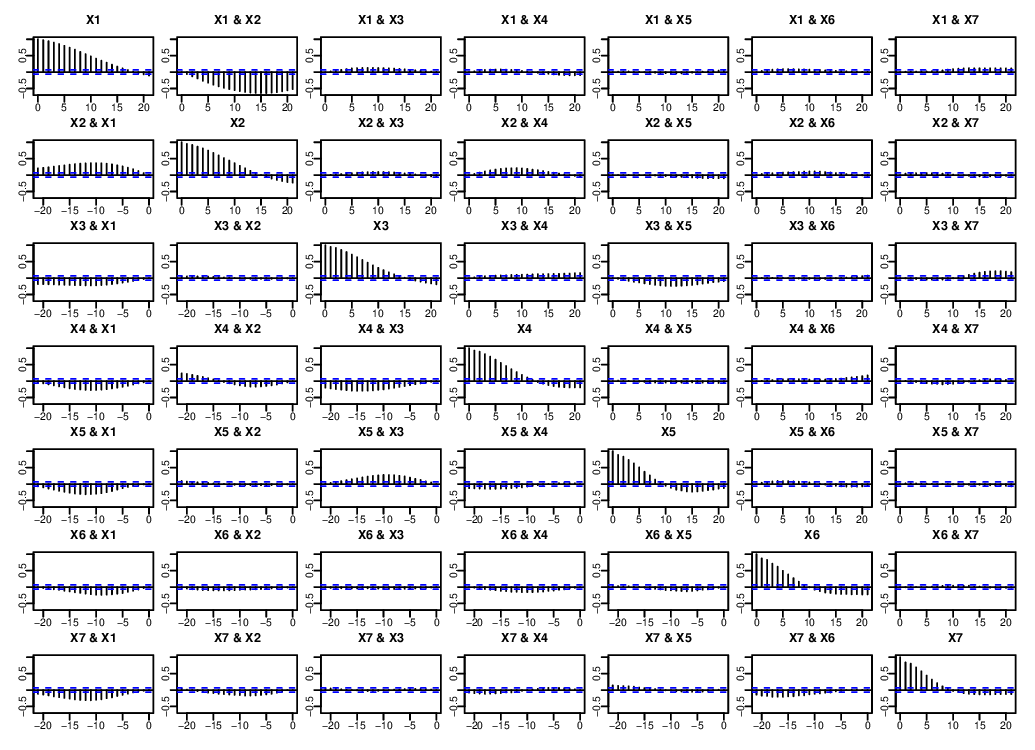}}\\
(b) Cross correlogram of the $7$ transformed component time series.
%\subfigure {\includegraphics[scale=0.363]{3exptong1000n10060}}\\
\setlength{\abovecaptionskip}{0pt}
\end{center}
\caption{Cross correlograms for Example \ref{ex-measles}.}\label{measlesACF}
\end{figure}

We apply prewhitening to each transformed component time series
% plotted in Fig.\ref{measlesTransf}
by fitting an AR
model with the order determined by AIC and with the maximum order set at
5.  Although all those 7 filtered time series behave like white noise,
there are still quite a  few small but significant cross
correlations here and there. Fig \ref{measlesRatio}(a) plots, in
descending
order, the maximum cross correlations $\wh L_n(i,j)$ defined in (\ref{b19})
for those 7 transformed and prewhitened series.
As $1.96/\sqrt{n}=0.064$ with  $n=937$ now, one may argue that the segmentation assumption
 does not hold for this example. Consequently the ratio
estimator $\wh r$ defined in (\ref{b14}) does not make any sense for this
example; see
also Fig \ref{measlesRatio}(b).

Nevertheless Fig \ref{measlesRatio}(a) ranks the pairs of
transformed component series according to the strength of the cross
correlation. If we would only accept $r$ connected pairs, this leads to
an {\sl approximate} segmentation according to the rule set in
Section \ref{sec220}. By doing this, we
effectively ignore some small, though still statistically
significant, cross correlations. Table\ref{tab1} lists the
different segmentations corresponding to the different values of
$r$.  It shows that the group \{4,\,5\} is always present until all
the 7 series merge together. Further it only takes 6 connected
pairs, corresponding to the 6 largest points in
Fig \ref{measlesRatio}(a), to merge all the series together.

\begin{figure}[tbh]
\begin{center}
\psfig{figure=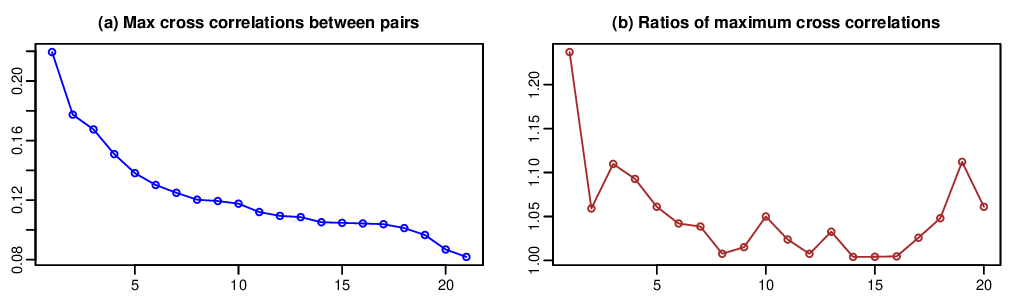, width=5in}
% \captionstyle{center}
\caption{{\rm(a)} The maximum cross correlations, plotted in descending
order,  among each of the $({7\atop 2} ) =21 $ pairs component
series of the transformed and prewhitened measles series. The
maximization was taken over the lags between -20 to 20. {\rm (b)} The
ratios of two successive correlations in (a).} \label{measlesRatio}
\end{center}
\end{figure}

\begin{table}
\begin{center}
%\captionstyle{center}
\caption{Segmentations determined by different
numbers of connected pairs for the transformed series in Example \ref{ex-measles}.} \label{tab1}
\begin{tabular}{c|c|l}
No. of connected pairs & No. of groups & \qquad Segmentation\\
\hline
1 & 6 & \{4,\,5\}, \{1\}, \{2\}, \{3\}, \{6\}, \{7\}\\
2& 5 & \{1,\,2\}, \{4,\,5\}, \{3\}, \{6\}, \{7\}\\
3& 4 & \{1,\,2,\,3\}, \{4,\,5\}, \{6\}, \{7\}\\
4& 3 & \{1,\,2,\,3,\,7\}, \{4,\,5\}, \{6\}\\
5& 2 & \{1,\,2,\,3,\,6,\,7\}, \{4,\,5\}\\
6& 1 & \{$1,\ldots,7$\}\\
\end{tabular}
\end{center}
\end{table}

The forecasting comparison is conducted in the same manner as in Examples \ref{ex-temperature} and \ref{ex-UNindustry}. We adopt the segmentation with 4 groups: \{1,\,2,\,3\}, \{4,\,5\}, \{6\} and \{7\}, i.e. we regard that only  the
three pairs, corresponding to the 3 maximum cross correlations in
Fig \ref{measlesRatio}(a), are connected.
 We forecast the notified measles cases in the
last 14 weeks of the period for all the 7 cities. Due to the fact that the
data from different cities are on different scales, we present the results based on relative MSEs in Table\ref{table7cities}. More specifically, we first fit an AR model for each component series of the original time series and calculate the associated MSEs. For a given other method, we define its relative MSE for each component series of the original time series as the ratio of its MSE and that of the fitted univariate AR model mentioned before. Once again the
forecasting based on this (approximate) segmentation is much more
accurate than those based on the direct VAR and RVAR models, and univariate AR model
to each of the original time series, although we have ignored quite
a few small but significant cross correlations among the transformed
series. The over-segmentation case with each
component as an individual group and the incomplete-segmentation case with 3
groups (\{1,\,2,\,3,\,7\}, \{4,\,5\},\{6\}) are also considered. The over-segmentation ignores all the correlations between any different components of the transformed series.
Such cross correlations in this example are very significant. Hence, the over-segmentation will
have an adverse effect while the incomplete-segmentation taking
account of more correlations will have an advantage in this case, which is also verified by the results presented in Table\ref{table7cities}.

\end{example}

\begin{example}
\label{ex-clothing} Now we consider the daily log-sales
of a clothing brand in 25 provinces in China in 1 January 2008 -- 9
December 2012 (i.e. $n=1805$ and $p=25$).
%  The sales in the natural logarithm scale in 8
% provinces (i.e. Beijing, Fujian, Guangdong, Guangxi, Hainan,
% Hebei, Henan and Hubei) are plotted in Fig.\ref{clothingplot}.
All those series exhibit peaks before the Spring Festival (i.e. the
Chinese New Year, typically around February). The cross correlogram
of the 8 randomly selected component series in Fig \ref{clothingacf}
indicates the strong cross correlations over different time lags
among the sales over different provinces. The strong periodic
components with the period 7 indicate a regular sales pattern over 7
different weekdays. By applying the proposed segmentation transformation and the permutation
based on the maximum cross correlations with $m=25$ in (\ref{b19}), the
transformed 25 time series are divided into 24 group with
only non-single-element group containing the 15th and the 16th
transformed series. The same grouping is obtained for $m$ between 14
and 30. Note for this example, we should not use small $m$ as the
autocorrelations of the original data decay slowly; see
Fig \ref{clothingacf}.
\begin{figure}[thb]
\begin{center}
\psfig{figure=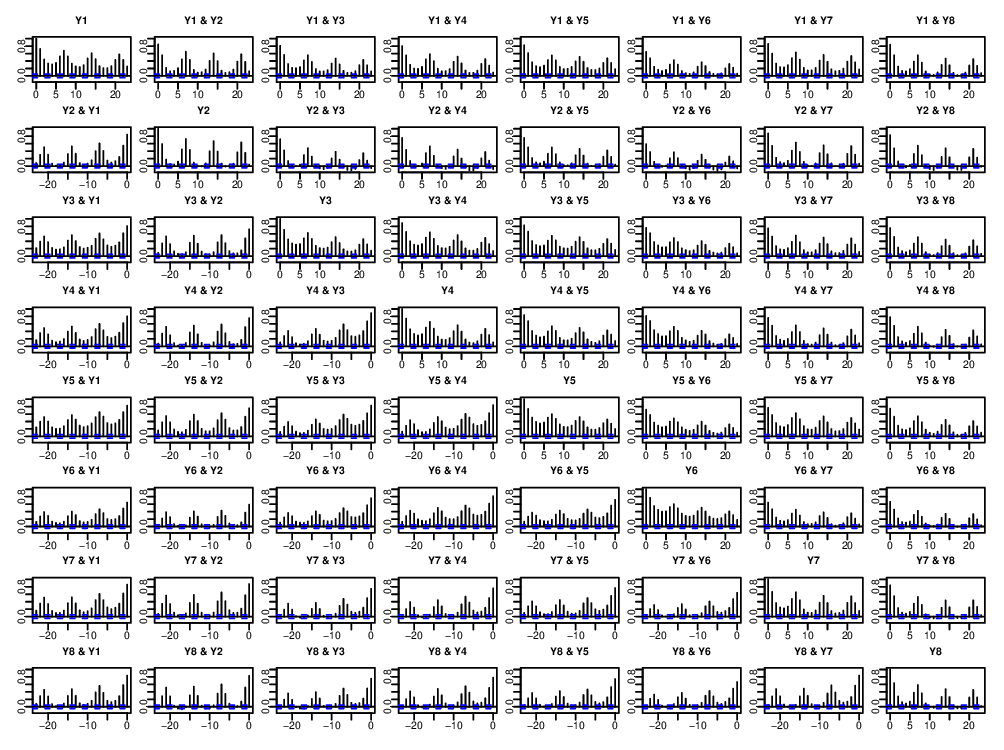, width=5in}
% \captionstyle{center}
 \caption{Cross correlogram of
the log-sales series of 8 randomly selected provinces in Example \ref{ex-clothing}.}\label{clothingacf}
\end{center}
 \end{figure}

To compare the post-sample forecasting performance, we calculate one-step-ahead
and two-step-ahead forecasts for each of the daily log-sales in the
last two weeks of the period. Table\ref{table7cities} list the means and the standard deviations
of the MSEs across the 25 provinces. With $p=25$, the fitted VAR(2)
model, selected by AIC, contain $2\times 25\times 25=1250$
parameters, leading to poor post-sample forecasting. The
RVAR(2) model improves the forecasting a bit, but it is still significantly
worse than the forecasting based on the approach of fitting a
univariate AR model to each of the original series directly. Since the
proposed segmentation leads to 24 subseries, it also fits univariate
AR models to 23 (out of 25) transformed series, fits a
2-dimensional VAR model to the 15th and the 16th transformed
series together. The proposed approach leads to much more accurate
forecasts as both the mean and standard deviation are much smaller
than those of the other three methods. The above comparison shows clearly that the cross correlations in
the sales over different provinces are valuable information which
can improve the forecasting for the future sales significantly.
However the endeavour to reduce the dimension by, for example, TS-PCA, is necessary in order to make use of
this valuable information. We also consider an over-segmentation by regarding each component of the transformed series as an individual group, and
an incomplete-segmentation with \{5,\,15,\,16\} as a group and the other $23$ components as
$23$ individual groups. Both of them have good performance.

\end{example}

\setcounter{equation}{0}

\section{Segmenting multivariate volatility processes}
\label{se:volatility}

The methodology proposed in Section \ref{sec2} can be readily extended to
segment multivariate volatility processes. To this end, let $\by_t$ be a
$p\times 1$ volatility process. Let $\calF_t = \sigma( \by_t, \by_{t-1},
\ldots)$ and $ \var(\by_t | \calF_{t-1} ) = \bSigma_y(t)$. Without
loss of generality, we assume $\mathbb{E}(\by_t| \calF_{t-1}) =\bzero$ and $\var(\by_t) =
\bI_p$. Suppose that there exists an orthogonal matrix $\bA$ for which
$\by_t = \bA \bx_t$ and
$
\var(\bx_t | \calF_{t-1} ) = \diag \{ \bSigma_1(t), \ldots, \bSigma_q(t) \}
$
with $\bSigma_1(t), \ldots, \bSigma_q(t)$ being, respectively, $p_1\times p_1, \ldots,$ $p_q \times p_q$ non-negative definite matrices. Hence the latent $p$-dimensional volatility process $\bx_t$ can be segmented into $q$ lower-dimensional processes, and there exist no {\sl conditional} cross correlations across those $q$ processes.

%To recover the hidden segmentation in volatility,
Let
$
\bW_y = \sum_{B \in \calB_{t-1}} [ \mathbb{E} \{ \by_t \by_t^\T \mathbb{I}(B)\} ]^2$ and $\bW_x = \sum_{B \in \calB_{t-1}} [ \mathbb{E} \{ \bx_t \bx_t^\T \mathbb{I}(B)\} ]^2,
$
where $\calB_{t-1}$ is a $\pi$-class and
the $\sigma$-field generated by
$\calB_{t-1}$ equals to $\calF_{t-1}$.
Since  it holds for any $B \in \calB_{t-1}$ that
$
\mathbb{E} \{ \bx_t \bx_t^\T \mathbb{I}(B)\} = \mathbb{E}\{\mathbb{I}(B) \mathbb{E}(\bx_t \bx_t^\T| \calF_{t-1})\} = \mathbb{E}[ \mathbb{I}(B) \diag \{ \bSigma_1(t), \ldots, \bSigma_q(t) \}]
$
is a block diagonal matrix, so is $\bW_x $. Now (\ref{b3}) still holds
for the newly defined $\bW_y $ and $\bW_x$. Thus $\bA$ can be
estimated exactly in the same manner as in Section \ref{sec21}. An estimator
for $\bW_y$ can be defined as
$
 \wh \bW_y = \sum_{B \in \calB} \sum_{k=1}^{k_0} \{ (n-k)^{-1}
 \sum_{t=k+1}^n \by_t \by_t^\T \mathbb{I}(\by_{t-k}\in B) \}^2,
$
where $\calB$ is a set with elements $\{ \bu \in \mathbb{R}^p: \|\bu\|_2\le \|\by_t\|_2 \}$ for $t=1, \ldots, n$. See \cite{FanWangYao08}. We illustrate this idea by a real data example.

\begin{figure}[tbh]
\begin{center}
\subfigure{\includegraphics[scale=0.75]{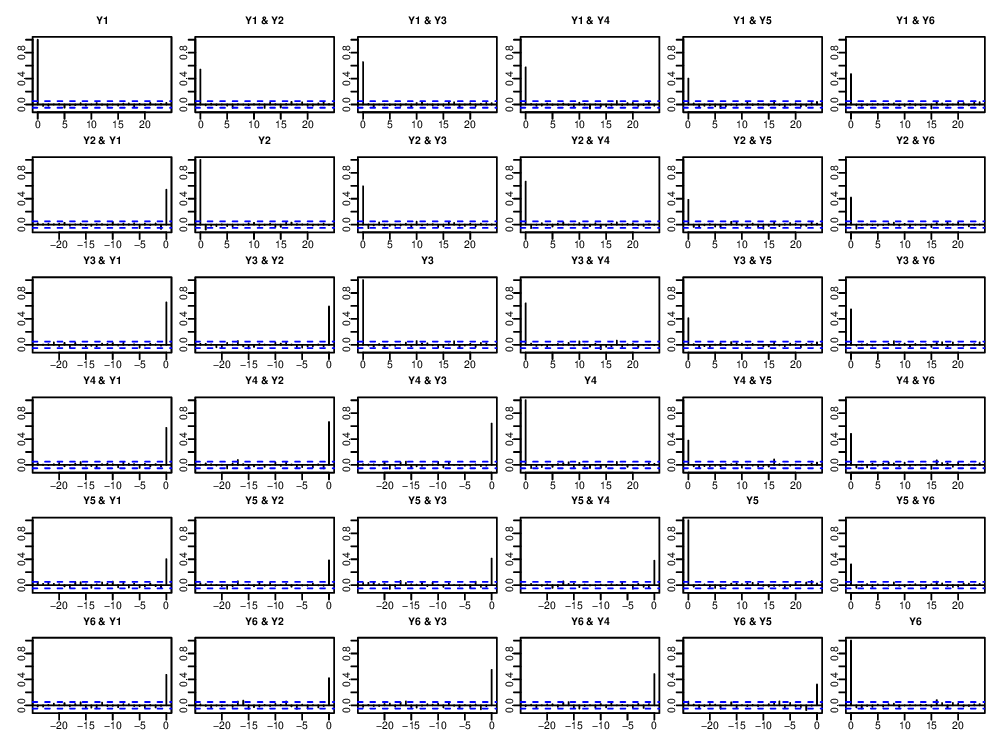}}\\
(a) Cross correlogram of
the residuals resulted from fitting each original component series a GARCH(1,1) model. \\
\subfigure{\includegraphics[scale=0.75]{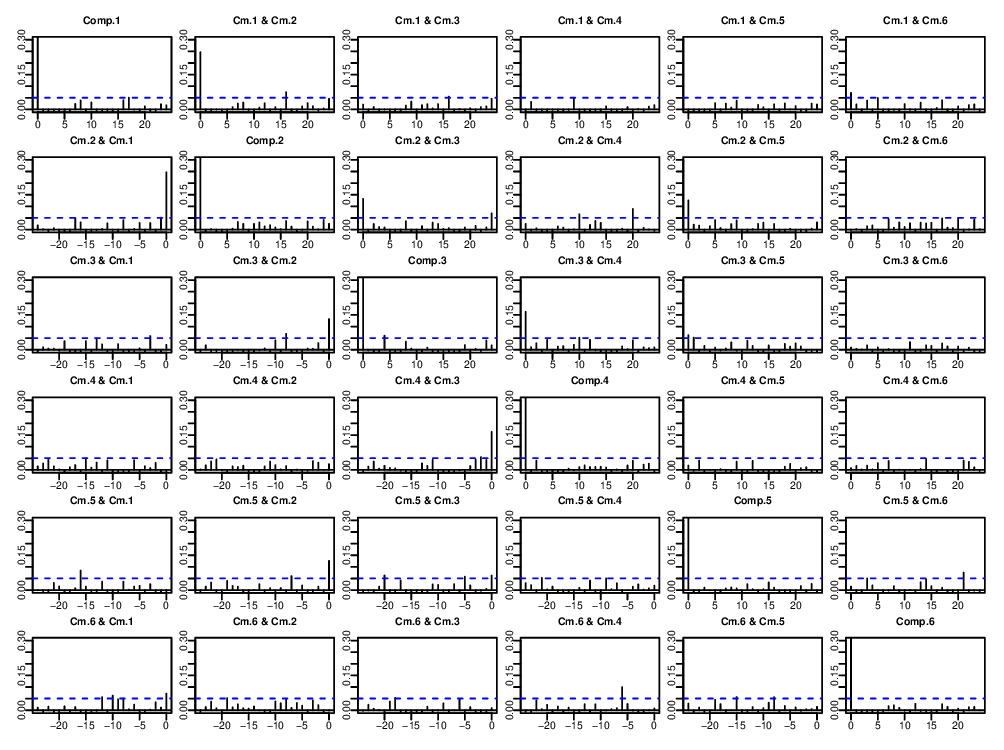}}\\
(b) Cross correlogram of
the  residuals resulted from fitting each series of PCA components a
GARCH(1,1) model.
\setlength{\abovecaptionskip}{0pt}
\end{center}
\caption{Cross correlograms for Example \ref{ex-volatility}.}
\label{return1}
\end{figure}

\begin{figure}[tbh]
\begin{center}
\psfig{figure=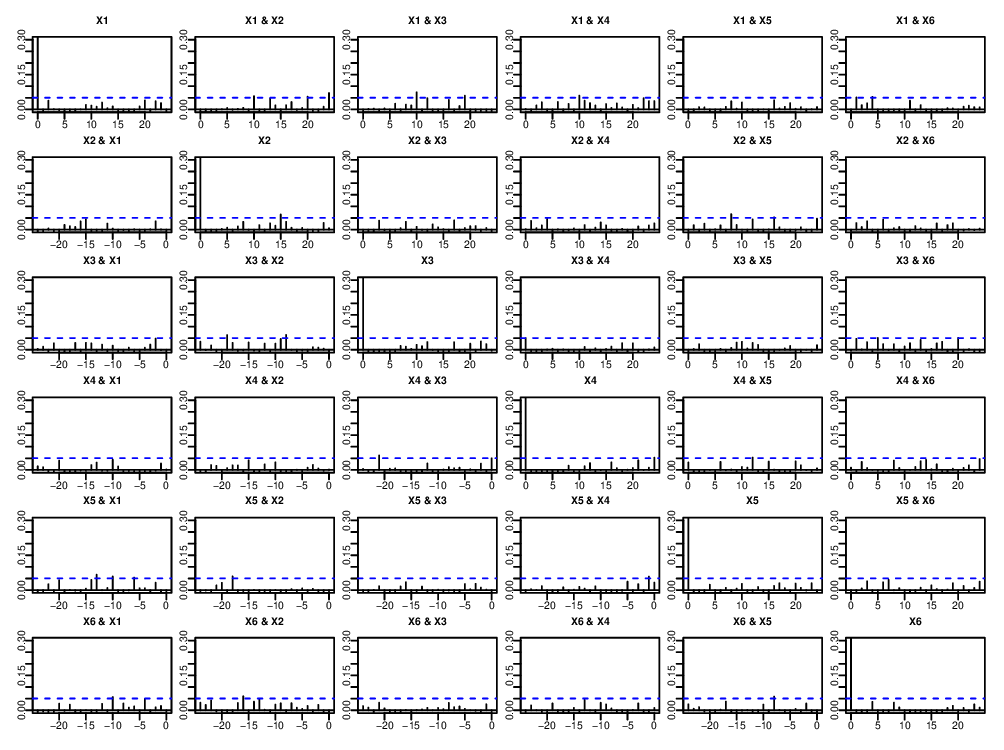, width=5in}
% \captionstyle{center}
 \caption{Cross correlogram of
the residuals resulted from fitting each component series of the transformed series $\wh
\bx_t$
with a GARCH(1,1) model in Example \ref{ex-volatility}. }
\label{returnTranspreACF}
\end{center}
\end{figure}

\begin{example} \label{ex-volatility}
We consider the daily returns of
the stocks of Walt Disney Company, Wells Fargo $\&$ Company, Honeywell
International Inc., MetLife Inc., H $\&$ R Block Inc. and Cognizant
Technology Solutions Corporation in 14 July 2008 -- 11 July 2014.
For this data set, $n=1509$ and $p=6$.
%of Bank of America Corporation, Dell Inc., JPMorgan Chase$\&$Co.,
%FedEx Corporation, McDonald's Corp. and American International Group
%in 2 January  2002 -- 10 July 2008. For this data, $n=1642$ and
%$p=6$.
% The cross correlogram of the squared daily returns of those 6 stocks
% is displayed in Figure \ref{returnsacf}, which shows clearly
% the significant autocorrelations and cross correlations.
Denote by $\by_t = (y_{1,t}, \ldots, y_{6,t})^\T$ the returns on
the $t$-th day. By fitting each return series a GARCH(1,1) model, we
calculate the residuals $\ve_{i,t} = y_{i,t}/\wh\sigma_{i,t}$ for
$i=1, \ldots, 6$, where $\wh\sigma_{i,t}$ denotes the predicted
volatility for the $i$-th return at time $t$ based on the fitted
GARCH(1,1) model.  The cross correlogram of the residual series
are plotted in Fig \ref{return1}(a), which shows the strong and
significant concurrent correlations across all residual series.
It indicates clearly that $\var(\by_t|\calF_{t-1})$ is not a block
diagonal matrix. We also apply the traditional PCA to
the 6 returns series, the cross correlogram of pre-whitened series
is shown in Fig \ref{return1}(b). There are also strong and significant
concurrent correlations across the residual series, see Panels
(1,\,2), (2,\,3), (3,\,4), (2,\,5) and (6,\,4). This indicates all the
principal components should not be modelled separately. Now we apply the segmentation transform stated above. We repeat the whitening process above for the transformed series
$\wh\bx_t$,
 i.e. fit an GARCH(1,1) model for each of the component series of
$\wh\bx_t$ and calculate the residuals. Fig \ref{returnTranspreACF} presents the cross
correlogram of these new residual series. There exist almost no
significant cross correlations among the residual series. This is
the significant evidence to support the assertion that
$\var(\bx_t|\calF_{t-1})$ is a diagonal matrix. For this example,
the segmentation method leads to the conditional uncorrelated
components of \cite{FanWangYao08}.

\end{example}

\section{Final remarks}\label{se:finalremark}

This paper proposes a contemporaneous linear transformation to
segment a multiple time series into several both contemporaneously
and serially uncorrelated subseries. The method is simple, and can be used as
a preliminary step to reduce a high-dimensional time series modelling problem
into several lower-dimensional problems. The reduction of dimensionality
is often substantial and effective.

The method is abbreviated as TS-PCA, as it can be viewed as a version of
PCA for multiple time series. Like the standard
PCA, TS-PCA technically also boils down to an eigenanalysis for a
positive definite matrix.  The difference is that the intended
segmentation is not guaranteed to exist. However one of the
strengths of the proposed TS-PCA is that
 even when the segmentation
 assumption is invalid,
it provides some approximate segmentations which
ignore some minor (though still significant) cross correlations and,
thus, lead to parsimonious modelling strategies. Those parsimonious
strategies often bring in improvements in, for example, forecasting future
values. See,
e.g., Example~\ref{ex-measles}.
Furthermore when the dimension of time
series is large,
TS-PCA is necessary in order to use the information across different
component series effectively. See, e.g., Example~\ref{ex-clothing}.

We have conducted some post-sample forecasting comparison with several
real data including some not reported in the paper. The forecasting based
on the proposed TS-PCA always outperforms that for the original data. We
give one explanation as follows. It follows from (\ref{b3}) that
$
\Omega \equiv
\tr(\bW_y) -p = \sum_{k=1}^{k_0} \sum_{i,j=1}^p \rho_{y,ij}^2(k) = \tr(\bW_x) -p
=\sum_{k=1}^{k_0} \sum_{i,j=1}^p \rho_{x,ij}^2(k),
%{\color{blue} = \sum_{k=1}^{k_0} \sum_i \rho^x_{ii}(k)^2 }
$
where $\rho_{y,ij}(k)$ and $\rho_{x,ij}(k)$ denote, respectively, the
cross correlation at lag $k$ between the $i$-th and the $j$-th components of $\by_t$ and $\bx_t$.
 Since the future prediction is based on the serial correlations,
$\Omega$ defined above can be taken as
 a measure for the predictive strength, which is the same for $\by_t$ and
$\bx_t$.  To make use the full predictive strength of $\by_t
$, we need to model the $p$-vector process appropriately to catch all the
autocorrelations and cross-correlations (over different time lags) among
the $p$ components of $\by_t$. In contrast, such a task for $\bx_t$ is
much easier as it can be divided into $q$ lower-dimensional problems. In
the ideal situation when $q=p$, i.e. $\rho_{x,ij}(k) =0$ for any $i\ne
j$, we just need to model all the component
series of $\bx_t$ {\sl separately} in order to make the full use of the
overall predictive strength.

\section*{Acknowledgements}

The authors sincerely thank
the Co-Editor, Associate Editor and three referees for their very constructive suggestions
and comments that led to substantial improvement of the paper.

%\begin{supplement}
%\sname{Supplement A}\label{suppA} \stitle{Title of the Supplement A}
%\slink[url]{http://www.e-publications.org/ims/support/dowload/imsart-ims.zip}
%\sdescription{Dum esset rex in accubitu suo, nardus mea dedit odorem
%suavitatis. Quoniam confortavit seras portarum tuarum, benedixit
%filiis tuis in te. Qui posuit fines tuos}
%\end{supplement}


\begin{thebibliography}{9}



\bibitem[{Anderson(1963)}]{Anderson_1963}
Anderson, T. W. (1963). The use of factor analysis in the statistical analysis of multiple time series. {\sl Psychometrika}, {\bf 28}, 1--25.


\bibitem[{Back and Weigend(1997)}]{BackWeigend_1997}
Back, A. D. and Weigend, A. S. (1997). A first application of independent component analysis to extracting structure from stock returns. {\sl Int. J. Neural Syst.}, {\bf 8}, 473--484.

\bibitem[{Bai and Ng(2002)}]{BaiNg_Econometrica_2002}
Bai, J. and Ng, S. (2002). Determining the number of factors in approximate factor models. \textsl{Econometrica}, 70, 191--221.

\bibitem[{Belouchrani et al.(1997)}]{Belouchrani_1997}
Belouchrani, A., Abed-Meraim, K., Cardoso, J.-F. and Moulines, E. (1997).
A blind source separation technique using second-order statistics.
{\sl IEEE T. Signal Proces.}, {\bf 45}, 434--444.

\bibitem[{Bickel and Levina(2008)}]{BickelLevina_2008}
Bickel, P. J. and Levina, E. (2008). Covariance regularization by thresholding. \AS, {\bf 36}, 2577--2604.

\bibitem[{Box and Jenkins(1970)}]{BoxJenkins_1970}
Box, G. E. P. and Jenkins, G. M. (1970). {\sl Time Series Analysis, Forecasting and Control.} Holden-Day, San Francisco.

\bibitem[{Box and Tiao(1977)}]{BoxTiao_1977}
Box, G. E. P. and Tiao, G. C. (1977). A canonical analysis of multiple time series. {\sl Biometrika}, {\bf 64}, 355--365.

\bibitem[{Brockwell and Davis(1996)}]{BrockwellDavis_1996}
Brockwell, P. J. and Davis, R. A. (1996). {\sl Introduction to Time Series and Forecasting}. Springer, New York.

\bibitem[{Brillinger(1981)}]{Brillinger_1981}
Brillinger, D. R. (1981). {\sl Time Series: Data Analysis and Theory}.
Holt, Rinehart and Winston, New York.


\bibitem[{Cardoso(1998)}]{Cardoso_1998}
Cardoso, J. (1998). Multidimensional independent component analysis.
{\sl Proceedings of the 1998 IEEE Int. Conf. Acoustics,
Speech and Signal Processing}, {\bf 4}, 1941--1944.

\bibitem[{Chang, Guo and Yao(2015)}]{ChangGuoYao_2013}
Chang, J., Guo, B. and Yao, Q. (2015). High dimensional stochastic regression with latent factors, endogeneity and nonlinearity. \JOE, {\bf189}, 297--312.

\bibitem[{Chang, Guo and Yao(2016)}]{CGY_supple}
Chang, J., Guo, B. and Yao, Q. (2016). Supplement to ``Principal component analysis for second-order stationary
vector time series."


\bibitem[{Davis, Zang and Zheng(2012)}]{Davis2012}
Davis, R. A., Zang, P. and Zheng, T. (2012). Sparse vector autoregressive modelling. Available at {\sl arXiv:1207.0520}.


\bibitem[{Fan, Wang and Yao(2008)}]{FanWangYao08}
Fan, J., Wang, M. and Yao, Q. (2008). Modelling multivariate volatilities via conditionally uncorrelated components. \JRSSB, {\bf 70}, 679--702.

\bibitem[{Fan and Yao(2003)}]{FanYao_2003}
Fan, J. and Yao, Q. (2003).
{\sl Nonlinear Time Series: Nonparametric and Parametric Methods}.
Springer, New York.

\bibitem[{Forni et al.(2005)}]{Fornietal_2005}
Forni, M., Hallin, M., Lippi, M. and Reichlin, L. (2005). The generalized dynamic factor model: One-sided estimation and forecasting. \JASA, {\bf100}, 830--840.

\bibitem[{Guo, Wang and Yao(2016)}]{GuoWangYao_2014}
Guo, S., Wang, Y. and Yao, Q. (2016). High-dimensional and banded vector autoregressions. \BKA, {\bf103}, 889--903.



\bibitem[{Han and Liu(2013)}]{HanLiu_2013}
Han, F. and Liu, H. (2013). A direct estimation of high dimensional stationary vector autoregressions. Available at {\sl arXiv:1307.0293}.

\bibitem[{Huang and Tsay(2014)}]{Huang_2014}
Huang, D. and Tsay, R. S. (2014). A refined scalar component approach to multivariate time series modeling. {\sl Manuscript}.

\bibitem[{Hyv{\"a}rinen, Karhunen and Oja(2001)}]{Hyvarinenetal_2001}
Hyv{\"a}rinen, A., Karhunen, J. and Oja, E. (2001). {\sl Independent Component Analysis}. Wiley, New York.

\bibitem[Jakeman, Steele and Young(1980)]{JakemanSteeleYoung_1980}
Jakeman, A. J., Steele, L. P. and Young, P. C. (1980). Instrumental variable algorithms for multiple input systems described by multiple transfer functions. {\sl IEEE T. Syst. Man. Cyb.}, {\bf 10}, 593--602.

\bibitem[{Lam, Yao and Bathia(2011)}]{LamYaoBathia_Biometrika_2011}
Lam, C., Yao, Q. and Bathia, N. (2011). Estimation of latent factors for high-dimensional time series. \BKA, {\bf98}, 901--918.

\bibitem[{Lam and Yao(2012)}]{LamYao_AOS_2012}
Lam, C. and Yao, Q. (2012). Factor modeling for high-dimensional time series: inference for the number of factors. \AS, {\bf40}, 694--726.

\bibitem[{Ledoit and Wolf(2004)}]{LedoitWolf_2004}
Ledoit, O. and Wolf, M. (2004). A well-conditioned estimator for
large-dimensional covariance matrices.
\JMA, {\bf 88}, 356--411.

\bibitem[{Liu, Xiao and Wu(2013)}]{LiuXiaoWu_2013}
Liu, W., Xiao, H. and Wu, W. B. (2013). Probability and moment inequalities under
dependence. \SS, {\bf23}, 1257--1272


\bibitem[{L\"utkepohl(2006)}]{Lutkepohl_(2006)}
L\"utkepohl, H. (2006). \textsl{New Introduction to Multiple Time Series Analysis}. Springer, Berlin.

\bibitem[{Matteson and Tsay(2011)}]{MattesonTsay_2011}
Matteson, D. S. and Tsay, R. S. (2011). Dynamic orthogonal components for multivariate time series. \JASA, {\bf 106}, 1450--1463.


\bibitem[{Pan and Yao(2008)}]{PanYao_2008}
Pan, J. and Yao, Q. (2008). Modelling multiple time series via common factors.
\BKA, {\bf 95}, 365--379.

\bibitem[{Paparoditis and Politis(2012)}]{PaparoditisPolitis_2012}
Paparoditis and Politis (2012). Nonlinear spectral density estimation: thresholding the correlogram. \JTSA, {\bf 33}, 386--397.

\bibitem[{Pe\~{n}a and Box(1987)}]{PenaBox_1987}
Pe\~{n}a, D. and Box, G. E. P. (1987). Identifying a simplifying structure in time series. \JASA, {\bf 82}, 836--843.



\bibitem[{Reinsel(1993)}]{Reinsel_1993}
Reinsel, G. C. (1993). {\sl Elements of Multivariate Time Series Analysis} (2nd edition). Springer.


\bibitem[{Rio(2000)}]{Rio_2000}
Rio, E. (2000). \textsl{Th\'{e}orie asymptotique des processus al\'{e}atoires faiblement d\'{e}pendants}. Springer, Berlin.



\bibitem[{Sarkar and Chang(1997)}]{SarkarChang_1997}
Sarkar, S. K. and Chang, C.-K. (1997). The Simes method for multiple hypothesis testing with positively dependent test statistics. \JASA, {\bf 92}, 1601--1608.

\bibitem[{Shojaie and Michailidis(2010)}]{ShojaieMichailidis_2010}
Shojaie, A. and Michailidis, G. (2010). Discovering graphical Granger causality using the truncated lasso penalty. \textsl{Bioinformatics}, {\bf26}, 517--523.

\bibitem[{Simes(1986)}]{Simes_1986}
Simes, R. J. (1986). An improved Bonferroni procedure for multiple tests of significance. {\sl Biometrika}, {\bf 73}, 751--754.

\bibitem[{Song and Bickel(2011)}]{SongBickel_2011}
Song, S. and Bickel, P. J. (2011). Large vector auto regressions. Available at {\sl arXiv:1106.3519}.


\bibitem[{Stewart and Sun(1990)}]{StewartSun_1990}
 Stewart, G. W. and Sun,
J. (1990). {\sl Matrix Perturbation Theory}. Academic Press.

\bibitem[{Stock and Watson(2002)}]{StockWatson_2002}
Stock, J. H. and Watson, M. W. (2002), Forecasting using principal components from a large number of predictors. \JASA, {\bf 97}, 1167--1179.

\bibitem[{Stock and Watson(2005)}]{StockWatson_2005}
Stock, J. H. and Watson, M. W. (2005). Implications of dynamic factor models for VAR analysis. Available at {\sl www.nber.org/papers/w11467}.

\bibitem[{Theis, Meyer-Baese and Lang(2004)}]{Theis_2004}
Theis, F. J., Meyer-Baese, A.  and Lang, E. W. (2004). Second-order blind source separation based on multi-dimensional autocovariances. In {\sl Independent Component Analysis and Blind Signal Separation} (Edi. C.G. Puntonet and A. Prieto). Springer, 726-733.


\bibitem[{Tiao and Tsay(1989)}]{TiaoTsay_1989}
Tiao, G. C. and Tsay, R. S. (1989). Model specification in multivariate time series (with discussion). \JRSSB, {\bf51}, 157--213.

\bibitem[{Tong, Xu and Kailath(1994)}]{Tong_1994}
Tong, L. Xu, G. and Kailath, T. (1994). Blind identification and
equalization based on second-order statistics: a time domain approach.
{\sl IEEE T. Inform. Theory}, {\bf 40}, 340--349.

 \bibitem[{Tsay(2014)}]{Tsay_2014}
 Tsay, R. (2014). {\sl Multivariate Time Series Analysis}. Wiley.



\end{thebibliography}
\end{document}